\documentclass[aip,amsmath,amssymb,reprint]{revtex4-1}
\usepackage{graphicx}
\usepackage{dcolumn}
\usepackage{bm}
\usepackage[utf8]{inputenc}
\usepackage[T1]{fontenc}
\usepackage{mathptmx}

\usepackage{subfigure}
\usepackage{paralist}
\usepackage{xcolor}
\usepackage{bbold}
\usepackage[inline]{enumitem}

\DeclareSymbolFont{newfont}{OML}{cmm}{m}{it}
\DeclareMathSymbol{\epsilon}{3}{newfont}{15} 
\DeclareMathSymbol{\varrho}{3}{newfont}{37} 

\def \beq {\begin{eqnarray}}
\def \eeq {\end{eqnarray}}

\def \Lowdin {{L\"{o}wdin }}

\newcommand{\bra}{\langle}
\newcommand{\ket}{\rangle}

\begin{document}
\title{Energy-weighted density matrix embedding of open correlated chemical fragments}
\author{Edoardo~Fertitta}
\author{George~H.~Booth}
\email{george.booth@kcl.ac.uk}
\affiliation{Department of Physics, King's College London, Strand, London, WC2R 2LS, U.K.}

\date{\today}

\begin{abstract}
We present a multi-scale approach to efficiently embed an {\em ab initio} correlated chemical fragment described by its energy-weighted density matrices, and entangled with a wider mean-field many-electron system. This approach, first presented in {\em Phys. Rev. B}, {\bf 98}, 235132 (2018), is here extended to account for realistic long-range interactions and broken symmetry states. The scheme allows for a systematically improvable description in the range of correlated fluctuations out of the fragment into the system, via a self-consistent optimization of a coupled auxiliary mean-field system. It is discussed that the method has rigorous limits equivalent to existing quantum embedding approaches of both dynamical mean-field theory, as well as density matrix embedding theory, to which this method is compared, and the importance of these correlated fluctuations is demonstrated. We derive a self-consistent local energy functional within the scheme, and demonstrate the approach for Hydrogen rings, where quantitative accuracy is achieved despite only a single atom being explicitly treated.
\end{abstract}

\maketitle

\section{Introduction} \label{sec:intro}

Density functional theory and mean-field approaches are generally the default choice for probing large correlated many-electron problems, due to their favorable scaling with system size. 
However, for strongly correlated problems (often involving transition metal atoms, structures away from equilibrium, or magnetic effects amongst others) the situation is more complicated as these strong correlation effects of electrons in narrow-band, localized orbitals result in dominating electron-electron repulsion where the standard mean-field and exchange-correlation functionals often fail qualitatively\cite{doi:10.1021/cr200107z}. Levels of theory beyond DFT are then necessary. Fragmentation approaches, where the localized orbitals from which these effects manifest are treated at a more accurate level of theory (e.g. quantum chemical wave function methods), offer a promising route forwards to avoid the approximations of density-functionals, but which allow for embedding in a wider mean-field to retain the favourable scaling with respect to the system size\cite{doi:10.1021/acs.accounts.6b00356}.

The embedding of a selected (generally small) set of strongly correlated degrees of freedom in a wider mean-field description of a system has become popular in recent years, and fall under the umbrella term of `quantum embedding' or `quantum cluster' methods\cite{cluster1,cluster2,cluster3,PhysRevLett.109.186404,Gull07,PhysRevB.90.085102,PhysRevB.96.085139,PhysRevB.95.045103}, which have recently been shown to be very succesful in describing lattice models \cite{PhysRevB.97.075112,Fan1951,Zheng2016}, realistic materials\cite{Park2014,Leonov2015}, as well as quantum chemistry\cite{Wouters2016,Weber5790,Chen2015,PhysRevLett.106.096402}.
The key difference to many other types of locally correlated, fragmentation or embedding approaches is that these chosen correlated degrees of freedom (commonly called `fragment' or `impurity' orbitals in the literature) must be treated as a truly {\em open} subsystem in the quantum mechanical sense, generally within an overall stationary state. This contrasts with many other local correlation methods, which start from the requirement of localized, fully occupied orbital states of some initial single-particle hamiltonian, which then describe fragments with no single-particle-like fluctuations from them or entanglement with the wider system.

Given instead an arbitrary choice of these fragment orbitals to describe physical correlated regions, they must be allowed to describe a {\em mixed} quantum state, where there is not just particle and energy exchange with the surrounding environment, but also the quantum mechanical fluctuations required to express the entanglement of the fragment with their environment. These intrinsically quantum fluctuations are incredibly important for stabilizing the fragment, and to ensure that quantum numbers which are normally conserved for a pure quantum state (such as spin or electron number) are generally no longer good quantum numbers within the fragment. This is physical, as we do not expect an arbitrary fragment of a quantum many-body system to retain good quantum numbers (such as electron number), indicating the presence of spin and/or charge quantum fluctuations from the fragment into its environment.

In order to describe this mixed-state quantum fragment, it should be quantified by a quantum variable other than a wavefunction, as this must by construction instead be used to describe a pure state. Dynamical mean-field theory (DMFT) has become popular in the last couple of decades\cite{cluster2,PhysRevLett.62.324,cluster3,Zgid2011,PhysRevLett.106.096402}, where the fragment state is characterized by its one-particle Green function. This allows for these virtual quantum fluctuations from the fragment into the environment due to the correlated behavior of the fragment electrons. These fluctuations are considered explicitly, as a manifestly time-dependent problem of electron or hole propagation from the system into the environment and back over all time-scales, casting the environment entanglement as temporal fluctuations. Working with the correlated effects on this one-particle propagator allows a description of the physics arising from single-particle quantum fluctuations into the environment from the fragment, including delocalization of charge or spin density into the environment, and allowing for a description of covalent bonds that the fragment electrons participate in. Added to this, all of the {\em local} correlated interactions of the fragment electrons are explicitly included in the physics of the fragment Green function, describing the effects of these strong Coulomb repulsion arising from these local orbitals. While non-local many-body interactions between the fragment and environment are truncated, through the self-consistency this single-particle propagator can describe the single-particle fluctuations in a true correlated sense, leading to a fully correlated description of the system-environment coupling.

However, the computational complexities and costs associated with the description of explicit time-dependent processes (despite a time-independent hamiltonian) can be significant, and because of this, the fully static `density matrix embedding theory' (DMET) was developed\cite{PhysRevLett.109.186404,doi:10.1021/ct301044e,Wouters2016}. In this, the fragment is described by its (static) one-particle density matrix, which is self-consistently optimized with the inclusion of the local correlated (two-particle) interactions. While this one-body density matrix of the fragment can describe a mixed state, this description of the fragment precludes the ability to explicitly characterize the fragment-environment fluctuations at any lengthscale arising directly from the correlated physics. This explicit information is simply not contained within the fragment-space reduced one-particle density matrix.
More specifically, we can consider a diagonal element of the fragment one-particle density matrix as a probability amplitude associated with an electron at a position ${\bf r_i}$. From a path integral perspective, this can be computed from the propagator arising from a sum of all paths which start and end at ${\bf r_i}$, including those that leave the fragment space, propagate through the environment, and return. The inclusion of the correlated physics of the fragment can dramatically change the weights of these paths, modifying in turn the fluctuations into the environment, and properties of the fragment space. In DMET, it is only the correlation-driven changes to the entirely local part of these paths which is self-consistently matched at each iteration (the reduced density matrices), while physical effects arising due to the coupling of the correlated fragment to its environment and its quantum fluctuations are simply approximated via the mean-field rather than explicitly described and optimized. As a potentially more severe and further uncontrolled constraint, these reduced fragment descriptors can only be matched via an idempotent description of the full system, ruling out a general matching of the fragment and mean-field full system density matrices\cite{Tsuchimochi2015,PhysRevB.96.235139}. Despite these limitations, the DMET approach has proven successful in many settings, where the physics is relatively local such that sufficiently large fragment spaces can be found to mitigate these approximations, made possible by its relatively computationally efficient formulation\cite{Zheng1155,PhysRevX.5.041041,PhysRevB.93.035126}.

Very recently, the authors presented an alternative approach, which formally overcomes the limitations of the DMET method, dubbed `Energy-weighted Density Matrix Embedding Theory' (EwDMET)\cite{Fertitta2018}. This allows for a systematically improvable and self-consistent description of the missing correlated (one-particle) quantum fluctuations from the fragment into the environment, in a rigorously self-consistent fashion. This was also achieved whilst retaining the computational simplicity of a static, efficient method. Instead of describing the correlated fragment solely by the one-particle reduced density matrix, it is instead described via the {\em energy-weighted} one-particle density matrices (EwRDMs). The zeroth order energy-weighted density matrix is simply the traditional one-particle density matrix, while the self-consistent matching of the higher-order variants describe increasingly long-range correlated changes to the local fragment expectation values, quantifying the beyond-mean-field quantum fluctuations into the environment.

These variables therefore characterize a length scale of the Feynman paths into the environment which are allowed to contribute to local, fragment one-particle correlated expectation values. Increasing this length scale allows for an increasingly well-resolved, rigorously self-consistent, and manifestly many-body description of these quantum fluctuations from the fragment, and a seamless integration of the fragment into the wider system. Increasing this lengthscale to convergence formally provides the same physical description of the system as DMFT, albeit in a fully time-independent formulation. Therefore the method can interpolate between the DMET and DMFT levels of theory. These EwRDMs of the fragment are found at each iteration via an exact (FCI) solution of an reduced-dimensionality `embedded' hamiltonian, found via an analytic projection from the full system hamiltonian. This embedding avoids any numerical fit, and is defined by its ability to exactly match the fragment EwRDMs from the full system.

Furthermore, the EwDMET approach allows for a rigorous self-consistency to be formulated, such that the correlated energy-weighted density matrices of the embedded system can be formally matched to those of the full system at convergence. Formally matching correlated and mean-field descriptions of these one-electron quantities is generally achieved via consideration of a time-dependent one-particle self-energy (as in DMFT). Retaining only a static potential, as is done in DMET, renders this generally impossible, and as a consequence DMET is only able to minimize the normed distance between these quantities\cite{Tsuchimochi2015,PhysRevB.96.235139}. However, the effects of a dynamical self-energy can also be reproduced, by instead considering the full system itself as an open system, connected to further fictitious degrees of freedom, via a static one-particle coupling. In the interests of clarity in the nomenclature, we shall denote the degrees of freedom of this extended, fictitious environment as {\em `auxiliary'} orbitals (to distinguish them from the `bath' orbitals of the cluster problem of section~\ref{section:cluster_ham}). These auxiliary orbitals hybridize with the fragment degrees of freedom, such that the mean-field description of this physical and auxiliary supersystem results in true correlated energy-weighted density matrices of the physical fragment space (and any symmetrically equivalent spaces), once this fictitious auxiliary space is traced out. This process allows for a self-consistent description of these correlated degrees of freedom, and their quantum fluctuations into their environment.

In a previous paper\cite{Fertitta2018}, this approach was demonstrated for a number of correlated lattice models, and found to provide excellent results with respect to various benchmarks. In this work, we extend the approach to investigate its viability for realistic {\em ab initio} molecular systems, where a long-range Coulomb interaction is included. For these initial studies of open, highly correlated subsystems, we choose to consider the paradigmatic model of stretching Hydrogen chains. The embedding allows for a mean-field scaling, and so large 50 atom rings can be treated, while we test the accuracy of just a single atomic fragment as the embedded region within a minimal basis. This system allows for a controllable model of weaker correlation (at close to equilibrium geometries), and strong correlation (at stretched geometries), and for which there already exists accurate benchmarks due to the one-dimensional topology via the density matrix renormalization group (DMRG)\cite{DMRG,DMRG2,DMRG3}. 

While describing the theory in section \ref{section:ewdmet}-\ref{section:algorithm}, we focus on quantum chemical applications and  extend the formal derivation to treat multiple fragment orbitals, as well as spin-symmetry broken mean-fields, allowing for spontaneously polarized fragments. The results of section \ref{section:results} show that allowing for this flexibility improves the results for these systems in the difficult recoupling region of bond lengths. In addition, an appropriate energy functional is derived for the method in section \ref{section:energy}, which can be obtained directly from the zeroth- and first-order energy-weighted density matrices. These energies are found to compare very well to benchmark values from DMRG. We also discuss the physical character of the bath orbitals used to solve for the energy-weighted density matrices of the fragment problem, and analyze the convergence with respect to the maximum order of these self-consistent EwRDMs in the method, as longer-ranged fluctuations from the fragment are explicitly optimized.

\section{Energy weighted density matrices}\label{section:ewdmet}

In this work, open correlated fragments of the molecule are described by their $n^{th}$ order energy-weighted reduced density matrices (EwRDMs), denoted $T^{(n)}_{h,\alpha \beta}$ and $T^{(n)}_{p,\alpha \beta}$ for the hole and particle sector respectively, where $\{\alpha, \beta \}$ label the spin-orbitals of the chosen molecular fragment. These quantities can be individually expressed in terms of expectation values of the ground state of the full system $| \Psi \rangle$, formed from nested commutators of a Hamiltonian ${\hat H}$\cite{doi:10.1063/1.4901432}, as
\begin{align}
    T_{h, \alpha \beta}^{(n)}&=\langle \Psi | c_{\beta}^{\dagger} \left[c_{\alpha}, \hat H \right]_{\{n\}} | \Psi \rangle \label{eqn:mom_h_commutators} \\
    T_{p, \alpha \beta}^{(n)}&=\langle \Psi | \left[c_{\alpha}, \hat H \right]_{\{n\}} c_{\beta}^{\dagger} | \Psi \rangle \label{eqn:mom_p_commutators},
\end{align}
where $\left[c_{\alpha}, \hat H \right]_{\{n\}} =  \left[ \dots [ [ c_{\alpha}, \hat H ], \hat H ], \dots \hat H \right]$ with $n$ total commutators and $\left[c_{\alpha}, \hat H \right]_{\{0\}} = c_{\alpha}$. These expressions reduce to
\begin{align}
T_{h, \alpha \beta}^{(n)}&=\langle \Psi | c_{\beta}^{\dagger} (E_0 - \hat H)^n c_{\alpha} | \Psi \rangle		\label{eqn:CorrHole}\\
T_{p, \alpha \beta}^{(n)} &= \langle \Psi | c_{\alpha} (\hat H - E_0)^n c_{\beta}^{\dagger} | \Psi \rangle,	\label{eqn:CorrPart}
\end{align}
where $E_0$ is the ground state energy. The zeroth-order hole EwRDM is simply the traditional one-body reduced density matrix, defined as
\begin{equation}
T_{h, \alpha \beta}^{(0)} = D_{\alpha \beta} = \langle \Psi | c_{\alpha}^{\dagger} c_{\beta} | \Psi \rangle	.	\label{eqn:1RDM}
\end{equation}
These expectation values can be defined in the $(N-1)$ and $(N+1)$ particle many-body eigenstates/values of the Hamiltonian $\{|\Psi_j^{(N-1)} \rangle; E_0 - \xi_j \}$ and $\{|\Psi_k^{(N+1)}\rangle; E_0 + \xi_k \}$. This gives the general form for these matrices as
\begin{align}
T_{h, \alpha \beta}^{(n)} &= \sum_j {\mathcal C}_{\alpha j} {\mathcal C}^*_{\beta j} (\xi_j - \mu)^n 	\label{eqn:eigmoms_h} \\
T_{p, \alpha \beta}^{(n)} &= \sum_k {\mathcal C}_{\alpha k} {\mathcal C}^*_{\beta k} (\xi_k - \mu)^n	\label{eqn:eigmoms_p},
\end{align}
where ${\mathcal C}_{\alpha j}=\langle \Psi | c^{\dagger}_{\alpha} | \Psi_j^{(N-1)} \rangle$ and ${\mathcal C}_{\alpha k}=\langle \Psi | c_{\alpha} | \Psi_k^{(N+1)} \rangle$ define the Dyson orbitals\cite{doi:10.1063/1.2805393} of the system of the hole and particle states respectively, $\mu$ is the chemical potential in a grand canonical ensemble, and $j$ and $k$ run over the entire set of eigenstates of the $N-1$ and $N+1$ sectors respectively.

If we instead restrict the full system to be defined by an (effective) single-particle hamiltonian, then the ground-state wavefunction can be found as a single Slater determinant. We can denote the eigenvectors/values of this single-particle Hamiltonian as $\{ {\bf C} ; \epsilon \}$, with $\xi_j=\epsilon_j$ as a consequence of Koopmans' theorem. In this case, the EwRDMs can be expressed from sums over occupied and virtual single-particle states respectively, as
\begin{align}
    \tilde T_{h, \alpha \beta}^{(n)} &= \sum_{\epsilon_j < \mu} C_{\alpha j} C_{\beta j}^{*}(\epsilon_j - \mu)^n	\label{eqn:latmom_h} \\	
    \tilde T_{p, \alpha \beta}^{(n)} &= \sum_{\epsilon_k > \mu} C_{\alpha k} C_{\beta k}^{*}(\epsilon_k - \mu)^n	\label{eqn:latmom_p}  ,
\end{align}
where the tilde simply denotes their construction from single-particle quantities. As it can be seen, there is no fundamental difference in formalism between the `correlated' quantities of Eqs.~\ref{eqn:eigmoms_h} and \ref{eqn:eigmoms_p}, and their `uncorrelated' counterparts in Eqs.~\ref{eqn:latmom_h} and \ref{eqn:latmom_p}, where ${\mathcal C}_{\alpha i}=C_{\alpha i}$. However, since the two-body interactions in the correlated Hamiltonian cause splittings of the hole and particle states, the number of many-body states which are summed over to construct the correlated EwRDMs is exponentially larger than in the single-particle case.

The aim of the approach described here (EwDMET)\cite{Fertitta2018}, is to self-consistently match the matrices ${\bf {\tilde T}}^{(n)}_{h/p}={\bf T}^{(n)}_{h/p}$ spanning a correlated fragment of the molecule, for all EwRDMs up to a given order, denoted by $n_{\rm mom}$. The quantities in Eqs.~\ref{eqn:CorrHole} and \ref{eqn:CorrPart} are solved by FCI\cite{pyscf} in a subspace hamiltonian which includes the explicit two-body interactions of the fragment as 
discussed in section~\ref{section:cluster_ham}. Through the self-consistency, the full system one-particle hamiltonian is updated to modify the quantities in Eqns.~\ref{eqn:latmom_h} and \ref{eqn:latmom_p} in order to match these correlated quantities. In the case that symmetry equivalent fragments exist, then these quantities are also used to describe all fragments related by an allowed symmetry operation and update the mean-field hamiltonian accordingly.

The `density matrix embedding theory' (DMET) becomes a limit of EwDMET at $n_{\rm mom}=0$, where no quantum fluctuations out of the fragment are explicitly matched.
Moreover, the self-consistency in DMET is achieved by means of a local static `correlation' potential over the fragment space, $v_c$. However, this local potential is insufficient to rigorously match these quantities\cite{Tsuchimochi2015,PhysRevB.96.235139}. This inability is made even starker for higher order EwRDMs as attempted within EwDMET. In contrast to this, EwDMET allows for a rigorous self-consistency. Since the number of terms in the expressions for the EwRDMs is different in the single-particle and many-body calculation of the EwRDMs, this exact matching is achieved via coupling to an additional fictitious auxiliary system, which can induce additional hole and particle states to sum over. This construction is discussed in \ref{section:auxiliary}, while the full self-consistent procedure is described in \ref{section:algorithm}. However, to consider its connection to other methods, and to get further insight into the role of these quantities, it is first useful to explicitly connect them to the single-particle propagator (Green function) of the system.

\subsection{Connection to the fragment Green function} \label{sec:GFs}

In the introduction, the EwRDMs were rationalized in terms of modifications to Feynman paths which start and end in the correlated fragment, and hence define its local properties via inclusion of fluctuations into the environment. Each fragment EwRDM of order $n$ can be rigorously related to the sum of the weight of all self-returning paths of length $n$ from the fragment space (where the length scale is determined by an application of the Hamiltonian)\cite{moments_paths}. Therefore, as higher-order EwRDMs are included in the description of the correlated fragment, increasingly long-range modifications to the fragment description (propagator) are made as a direct result of the correlated interactions of the fragment. We can formalize these connections to the single-particle propagator (Green function) of the system, which is a framework which allows all environmental fluctuations on all time- and length-scales. This digression will allow for a further appreciation of the physics captured by the EwRDMs in this method, as well as also allowing for a connection to the successful quantum embedding method, `dynamical mean-field theory' (DMFT)\cite{cluster1,cluster2,cluster3,PhysRevLett.109.186404}. 

The single-particle propagator or Green function characterizes the time-dependent response of a system due to the insertion and subsequent propagation of an additional particle or hole (for these purposes into the ground state of the system). This quantity can be defined over many domains, including real-frequency, Matsubara frequency and imaginary time. We review these briefly here to define pertinent expressions, all for equilibrium, zero-temperature systems. The real-frequency, time-ordered Green function can be given in a general form as
\begin{equation}
G(\omega)_{\alpha \beta} = \lim_{\eta \rightarrow 0^+} \sum_j \frac{{\mathcal C}_{\alpha j} {\mathcal C}^*_{\beta j}}{\omega - (\xi_j - \mu) + i \eta {\rm sgn}(\xi_j - \mu)}	,
\end{equation}
where $j$ runs over all $N\pm1$ states. The corresponding total spectral function, ${\rm Tr}[A(\omega)_{\alpha \beta}]$, can then be uniquely and bijectively defined from
\begin{align}
A(\omega)_{\alpha \beta} &= -\frac{1}{\pi} {\rm Im} [G^R(\omega)_{\alpha \beta}]	\label{eqn:Spectrum} \\
&= \sum_j {\mathcal C}_{\alpha j} {\mathcal C}^*_{\beta j} \delta(\omega-(\xi_j-\mu))	,	\label{eqn:SpecFunc}
\end{align}
where $G^R(\omega)_{\alpha \beta}$ denotes the retarded rather than time-ordered version of the Green function\cite{Wetter}.

In Matsubara frequency, the Green function can also be given by a high-energy expansion\cite{doi:10.1063/1.4901432}, as
\begin{equation}
G(i\omega)_{\alpha \beta} = \sum_{n \ge 0} (-1)^{(n+1)} \frac{\langle \Psi | \{ [ {\hat H}, c_{\alpha} ]_{\{n\}} , c_{\beta}^{\dagger} \} | \Psi \rangle}{(i\omega)^n}.	\label{eqn:MatGF}
\end{equation}
Using the expression in Eqs.~\ref{eqn:CorrHole} and \ref{eqn:CorrPart}, we can write this as
\begin{multline}
G(i\omega)_{\alpha \beta} = \sum_{n \ge 0} \left( -\frac{1}{(i\omega)} \right)^n \left( \langle \Psi | c_{\alpha} ({\hat H}-E_0)^n c^{\dagger}_{\beta} | \Psi \rangle \right. \\
+ \left. \langle \Psi | c_{\beta}^{\dagger} (E_0 - {\hat H})^n c_{\alpha} | \Psi \rangle \right)	\label{eqn:LaurentGF}
\end{multline}
\begin{multline}
G(i\omega)_{\alpha \beta} = \sum_{n \ge 0} \left( -\frac{1}{(i\omega)} \right)^n \sum_j {\mathcal C}_{\alpha j} {\mathcal C}^*_{\beta j} \left( \Theta(\xi_j - \mu) (\xi_j - \mu)^n \right. \\
 + \left. \Theta(\mu - \xi_j) (\xi_j - \mu)^n \right)	,
\end{multline}
where $\Theta$ is the Heaviside step function. In these expressions, the first terms correspond to sums over particle ($N+1$ or virtual) states, while the second term sums over hole ($N-1$ or occupied) states of the system.

Finally, we can also Fourier transform this representation into the imaginary-time domain, and define a Green function as
\begin{multline}
G(\tau)_{\alpha \beta} = - \sum_j {\mathcal C}_{\alpha j} {\mathcal C}^*_{\beta j} e^{-\tau (\xi_j-\mu)} [\Theta(\tau) \Theta(\xi_j - \mu) \\
- \Theta(-\tau) \Theta(\mu-\xi_j)]	.	\label{eqn:ImTimeGF}
\end{multline}
This definition also separates contributions to the Green function in terms of particle and hole states.

These expressions clarify many of the relationships of the single-particle propagator to the energy-weighted density matrices used to describe the correlated effects of the fragment. These static quantities characterize the {\em dynamical} character of the single-particle spectrum of the fragment, embedded in the system. By comparing Eq.~\ref{eqn:SpecFunc} and Eqs.~\ref{eqn:eigmoms_h} and \ref{eqn:eigmoms_p}, it is clear that each increased order of the energy-weighted density matrices characterizes an additional {\em moment} of the hole and particle distributions of the spectral function, as
\begin{eqnarray}
T_{h, \alpha \beta}^{(n)} &= \int_{-\infty}^{\mu} A(\omega)_{\alpha \beta} \omega^n d\omega  \label{eq:mom_distr_h}\\
T_{p, \alpha \beta}^{(n)} &= \int_{\mu}^{\infty} A(\omega)_{\alpha \beta} \omega^n d\omega . \label{eq:mom_distr_p}
\end{eqnarray}
These descriptors therefore define the integrated weight (${\bf T}^{(0)}$), mean (${\bf T}^{(1)}$), variance (${\bf T}^{(2)}$), skew (${\bf T}^{(3)}$), bimodal (${\bf T}^{(4)}$), and higher order character of the {\em separate} hole and particle parts of the local spectral function, as well as their spatial dependence among the different fragment orbitals. The analytic character of these functions ensures that this expansion is a convergent expansion towards to complete dynamical resolution of this spectral function, containing all the energy-dependent information of the single-particle propagator. We therefore also denote these EwRDM matrices as `moments' for brevity in this work, understanding them in the context described above for the separate particle and hole spectral distributions.

A different interpretation of these quantities can be obtained from the imaginary-time and Matsubara formalisms, where additional insight into the physical content of these EwRDMs can be found. From a comparison of Eq.~\ref{eqn:ImTimeGF} and Eqs.~\ref{eqn:eigmoms_h} and \ref{eqn:eigmoms_p}, we can assert that the moments are related to the derivatives of the imaginary-time Green function and hence Taylor expansion for short hole/particle propagation times, as
\begin{eqnarray}
{\bf T}_{h}^{(n)} &=& (-1)^{n} \left. \frac{d^n {\bf G(\tau)}}{d \tau^n}\right|_{\tau=0^-} \\
{\bf T}_{p}^{(n)} &=& (-1)^{(n+1)}\left. \frac{d^n {\bf G(\tau)}}{d \tau^n}\right|_{\tau=0^+},
\end{eqnarray}
and therefore define a systematically improvable timescale for the resolution of quantum fluctuations into the environment. In the Matsubara formalism, the Laurent expansion of the Green function in terms of inverse frequency in Eq.~\ref{eqn:LaurentGF} can be seen to depend on the sum of the hole and particle moments at each order. This dictates the shape of the high-frequency tail of the Green function.

Through these many connections to the full propagator, we can see these descriptors as a truncated expansion of the full dynamical behaviour of the system, describing quantum fluctuations from the fragment of increasing length or time scales. Although a unique description of the full dynamics of the single-particle propagator is only exactly defined in the \mbox{large-$n_{\rm mom}$} limit, important insights into the correlation of the system can be obtain even from low-moments expansions. It should be pointed out that, since we are dealing with the particle and hole moments separately, rather than with their sum in the full central moment, these are not high-energy expansions, despite being short-time expansions of the Green functions\cite{Potthoff98,Potthoff01}. This gives us confidence that it can describe strongly correlated fragments, where local atomic-like orbitals strongly influence the electronic structure of the system. Furthermore, the avoidance of these explicitly dynamical Green functions in the construction of the theory significantly simplifies the numerics, ruling out the need for discretization schemes, grids and other approximations, as well as admitting far more efficient ground-state electronic structure methods to solve the correlated problem within a rigorous mapping to the embedded system.

\section{Auxiliary space and extended Fock matrix}\label{section:auxiliary}

The aim of the EwDMET approach is to match the particle and hole `moments' of correlated wave function calculations over the fragment space (Eqs.~\ref{eqn:mom_h_commutators} and \ref{eqn:mom_p_commutators}), to the analogous expressions coming from a one-electron hamiltonian over the full system (Eqs.~\ref{eqn:latmom_h} and \ref{eqn:latmom_p}). The approach to do this, involves considering the full system itself as physically coupled via one-electron, tight-binding-like terms, to fictitious, auxiliary degrees of freedom\cite{PhysRevLett.109.186404,doi:10.1002/9781119129271.ch8}. These additional degrees of freedom can shift the one-electron eigenstates of the mean-field system, as well as introduce an increased number of these states. In this way, they can mimic the additional states arising via the correlated two-electron effects in Eqs.~\ref{eqn:eigmoms_h} and \ref{eqn:eigmoms_p}, to exactly match these quantities. This addition of external degrees of freedom to induce these changes to mean-field-like quantities is entirely equivalent to the introduction of a frequency- or time-dependent (dynamical) self-energy for the computation of Green's functions, and therefore can also be used to match arbitrary order moments. Also, as done in standard DMET, a local correlation potential $v_c$ is added on each symmetry equivalent fragment space to mimic the static part of the self-energy. 

In the following, we denote the number of auxiliary degrees of freedom which are coupled to each fragment of interest as $n_{\rm aux}$. As this number increases, there is the flexibility to arbitrarily accurately match any number of moments, $n_{\rm mom}$. In case there may exist symmetry-equivalent copies of the fragment degrees of freedom, the auxiliaries which are coupled to the fragment can also be replicated on other parts of the molecule. More formally, we can write the total one-electron hamiltonian for the physical $+$ auxiliary systems as a sum of the Fock matrix over the physical degrees of freedom, and couplings to auxiliary sites, as
\begin{equation}
{\hat h}_{\rm tot} = {\hat f} + {\hat h}_{\rm aux}	.	\label{eqn:totham}
\end{equation}
The physical Fock matrix is constructed as normal, with 
\begin{equation}
    f_{\alpha\beta}= t_{\alpha\beta} + \sum_{\gamma\delta} \left[ \left(\alpha\beta|\gamma\delta\right) -  \left(\alpha\delta|\gamma\beta\right) \right] D_{\gamma\delta}    ,\label{eq:fock}
\end{equation}
where ${\bf D}$ is the density matrix and $\hat t$ the nuclear and kinetic one-particle hamiltonian.
It should be pointed out that the basis employed in Eq.~\ref{eq:fock} refers to all physical orbitals, including the fragment as well as their environment, but excluding the auxiliary space.

The auxiliary hamiltonian, ${\hat h}_{\rm aux}$ consists of one-electron terms coupling all equivalent fragment spaces (indexed by $\lambda$) to their respective auxiliary spaces, as
\begin{equation}
{\hat h}_{\rm aux} = \sum_{\lambda} {\hat h}_{\rm aux}^{\rm frag(\lambda)}	,	\label{eqn:sumaux}
\end{equation}
where
\begin{equation}
\begin{split}
 {\hat h}_{\rm aux}^{\rm frag(\lambda)} =& \sum_{\alpha \beta \in \rm frag(\lambda)} \nu_{\alpha \beta}^{c} {\hat c}_{\alpha}^{\dagger} {\hat c}_{\beta} \\
&+\sum_{k}^{n_{\rm aux}} \varepsilon_{k} {\hat c}_{k}^{\dagger} {\hat c}_{k} + \sum_{k}^{n_{\rm aux}} \sum_{\alpha \in \rm frag(\lambda)}  \nu_{\alpha k} ({\hat c}^{\dagger}_{\alpha} {\hat c}_{k} + h.c.) ,
\end{split} 	\label{eqn:auxham}
 \end{equation}
 where $k$ denotes auxiliary degrees of freedom, orthogonal to any physical states.
This extension of the Fock matrix with the local correlation potential and the auxiliary degrees of freedom contained in ${\hat h}_{\rm aux}$ allows it to implicitly capture correlation-driven one-body quantum fluctuations which are neglected in Hartree-Fock theory. A diagonalization of this enlarged physical $+$ auxiliary system returns effectively correlated and non-idempotent moments, when projected back into the physical space. This projection amounts to ensuring the sums in Eqs.~\ref{eqn:latmom_h} and \ref{eqn:latmom_p} run over the eigenstates of ${\hat h}_{\rm tot}$, while the $\{ \alpha, \beta \}$ indices denote only physical system orbitals. In order to optimize these quantities, the parameters $\{ \nu_c, \varepsilon,\nu \}$ need to be varied to match the information obtained from the wave function of the correlated subspace hamiltonian, whose construction is described in the next section. It should be noted that it is possible for multiple solutions for the auxiliary Hamiltonian to be found which match the desired moments. Even in the $n_{\rm mom}\rightarrow \infty$ limit the self-consistency can in principle converge to multiple stationary points, in the same way that multiple solutions for `static'-mean-field theory exist within Hartree-Fock theory, or indeed solutions within dynamical mean-field theory.

Upon extension of the Fock matrix with the additional degrees of freedom, the total number of electrons also needs to be augmented to match the moments. In our previous work\cite{Fertitta2018} this was achieved by working in a grand canonical ensemble, with a fictitious (low) temperature, with the total electron filling smoothly changed via the chemical potential. However, in this work we work at zero temperature with a fixed integer filling of the orbitals. This filling is determined by ensuring that the mid-point of the HOMO-LUMO gap of the extended Fock matrix  $\mu_{\rm ext}$ (the total chemical potential), is constrained to equal the same definition for the chemical potential of the physical Fock matrix $\mu$. This in turn places a constraint on the auxiliary parameters $\{\varepsilon,\nu \}$, while $\nu_c$ is kept traceless. This ensures that through the self-consistency, the number of electrons in the physical system is not changed by the presence of the auxiliary degrees of freedom, hence avoiding variations in total charge in the physical system.

\section{Bath construction and cluster Hamiltonian}\label{section:cluster_ham}

While section~\ref{section:auxiliary} is concerned with the ability to manipulate the single-particle states of the full system in order to arbitrarily modify the fragment mean-field moments, we also require accurate fragment moments to self-consistently match these to. These are computed from a subspace Hamiltonian including explicit two-body interactions, which can be solved to high accuracy, to provide the correlated expectation values of Eqs.~\ref{eqn:mom_h_commutators} and \ref{eqn:mom_p_commutators}. In EwDMET, these are approximated via the solution to a reduced dimensionality Hamiltonian, where the physical fragment states are augmented with bath states, obtained via a projection from the full physical (non-fragment) and auxiliary system. This space of fragment $+$ bath orbitals, to be solved via high-level correlated ground-state calculations, is denoted the {\em cluster} space. The Hamiltonian in this space consists of the explicit (bare) two-electron terms only over the fragment degrees of freedom of the cluster.

The main concern is how these bath orbitals are chosen, and which Hamiltonian should be projected into the cluster. The criterion used for determining these bath states, is that they must be able to {\em exactly} reproduce the fragment moments of Eqs.~\ref{eqn:latmom_h} and \ref{eqn:latmom_p}, in the case where the Hamiltonian projected into the cluster is just ${\hat h_{\rm tot}}$, the mean-field Hamiltonian. In the DMET approach, only the zeroth-order hole moment (the RDM) is optimized, so ${\hat h_{\rm tot}}={\hat f}$, and the bath orbitals which fulfil this criteria are obtained from the Schmidt decomposition of the mean-field (many-electron) state which results from the diagonalization of ${\hat f}$\cite{Wouters2016,doi:10.1002/9781119129271.ch8,PhysRevB.89.035140}. In the case of EwDMET, this procedure is not sufficient to faithfully reproduce an arbitrary number of higher order moments, and therefore requires extension to ensure that a chosen number of these moments, for both particle and hole sectors, are spanned by the cluster Hamiltonian. This is physically justified, as the higher moments describe longer-ranged particle/hole fluctuations from the fragment, and therefore require the cluster to span a larger space which can capture these fluctuations.

We denote the orbital space corresponding to the bath as $\{|b_{x}\rangle\}$. To ensure that we can faithfully represent an arbitrary number of moments, we construct the bath space via the Schmidt decomposition of wave functions of the form $\hat h_{\rm tot}^m {\hat c}_{\alpha}^{(\dagger)} |\Phi \rangle$, for hole (particle) bath orbitals respectively.
In this, 
$|\Phi \rangle$ is the mean-field state resulting from $\hat h_{\rm tot}$ with a given chemical potential $\mu$. The decomposition of wave functions of this form give a single bath degree of freedom for each fragment spin-orbital ($\alpha$), order ($m$), and hole/particle character. The decomposition results in bath states of the form
\begin{eqnarray}
    |b_{\alpha,m}^{\rm hole} \rangle &=& \sum_{\kappa \notin {\rm frag}} \sum_{\epsilon_i<\mu} \frac{\epsilon_i^m C_{\alpha i} C_{\kappa i}^*}{{\sqrt{N_{\alpha, m}^{\rm hole}}}}| \kappa \rangle \label{eqn:schmidtmom_h} \\
    |b_{\alpha,m}^{\rm particle} \rangle &=& \sum_{\kappa \notin {\rm frag}} \sum_{\epsilon_i>\mu} \frac{\epsilon_i^m C_{\alpha i} C_{\kappa i}^*}{{\sqrt{N_{\alpha, m}^{\rm particle}}}}| \kappa \rangle \label{eqn:schmidtmom_p},
\end{eqnarray}
where $\kappa$ denotes degrees of freedom in the extended Fock matrix (with eigenvalues and vectors $\{ {\bf \epsilon} ; {\bf C} \}$) orthogonal to the fragment space. $N$ values are the normalization constants of the resulting orbitals. It should be noted that these bath orbitals
are not necessarily orthogonal, and so an orthogonalization scheme and removal of linear dependencies which arise is necessary. This can reduce the overall number of bath states, but ensures that the full fluctuation space is still spanned. Once these bath orbitals are found, projection operators into this space can be constructed, as
\begin{equation}
P_{\rm bath} = \sum_x |b_x \rangle \langle b_x |	,
\end{equation}
with similar expressions for projections into the fragment space ($P_{\rm frag}$) and full cluster space ($P_{\rm clust}=P_{\rm frag} \oplus P_{\rm bath}$).

\begin{table}[t!]
\centering 
\begin{tabular}{l l l} 
\hline\hline 
$n_{\rm mom}$ & $m_{\rm max}$ & $n_{\rm bath}/n_{\rm frag}$ \\ [0.5ex] 
\hline 
0 & 0 & 1 (DMET) \\ 
1 & 0 & 1 \\
2 & 1 & 3 \\
3 & 1 & 3 \\
4 & 2 & 5 \\ 
5 & 2 & 5 \\ [1ex]
\hline
\end{tabular}
\caption{Decomposition order and maximum bath space sizes. $n_{\rm mom}$ denotes the maximum order of fluctuation moment which is to be faithfully represented in the cluster space. $m_{\rm max}$ denotes the maximum order of wave function in Eqs.~\ref{eqn:schmidtmom_h} and \ref{eqn:schmidtmom_p} which must be decomposed, and $n_{\rm bath}/n_{\rm frag}$ denotes the maximum possible number of bath orbitals per fragment orbital in the cluster space (excluding reductions due to linear dependencies).}
\label{tab:bathspaces}
\end{table}

It is now necessary to determine the maximum order of wave functions ($m=m_{\rm max}$) which are required in order to faithfully represent a desired maximum moment order ($n_{\rm mom}$). This is important, since it will determine the total number of bath orbitals ($n_{\rm bath}$) in the cluster problem, and hence the feasibility of the accurate wave function calculation. The DMET approach considers $n_{\rm mom}=0$, which decomposes the wave function $m=0$. The particle and hole components of $m=0$ are linearly dependent and hence redundant, which results in only one bath orbital per fragment orbital. This bath orbital in DMET is identical to the one constructed above. For higher moments, the particle and hole bath states are no longer equivalent, and have to be considered separately. However, a standard rule of perturbation theory indicates that in order to know the moments up to an order $2n_{\rm mom}+1$, it is sufficient to span the wave functions up to an order $n_{\rm mom}$\cite{Wigner80}. This results in $m_{\rm max} = \left \lceil \frac{n_{\rm mom}-1}{2} \right \rceil$, and hence a total bath space dimensionality per fragment orbital for a given moment order is $2\left \lceil \frac{n_{\rm mom}-1}{2} \right \rceil + 1$. 

The number of bath orbitals required per fragment orbital in the cluster space is summarised in Table~\ref{tab:bathspaces}. The result is that the number of bath orbitals now grows linearly with both the number of fragment degrees of freedom, as well as the maximum lattice fluctuation moment which is to be found. Furthermore, it can be seen that the original DMET approach to cluster space construction can in fact span both the zeroth and first moment, although only the zeroth moment is self-consistently optimized in DMET. Finally, DMFT can be considered a self-consistent approach in the limit $n_{\rm mom}\rightarrow \infty$. In this limit, the number of bath orbitals goes to infinity, which again is consistent with the DMFT approach, where for exact diagonalization and wave function solvers of the cluster problem, an infinite number of bath states are formally required to represent the continuous hybridization function\cite{PhysRevB.89.035148,Liebsch2012}. From another perspective, this expansion of bath orbitals denotes a systematic enlargement of the local occupied/virtual space spanned by the cluster, with the constraint that only canonical orbitals with a non-zero overlap with the fragment space can be accessed.

\begin{figure*}[t!]
\includegraphics[scale=0.65,trim={0 0 0 0},clip]{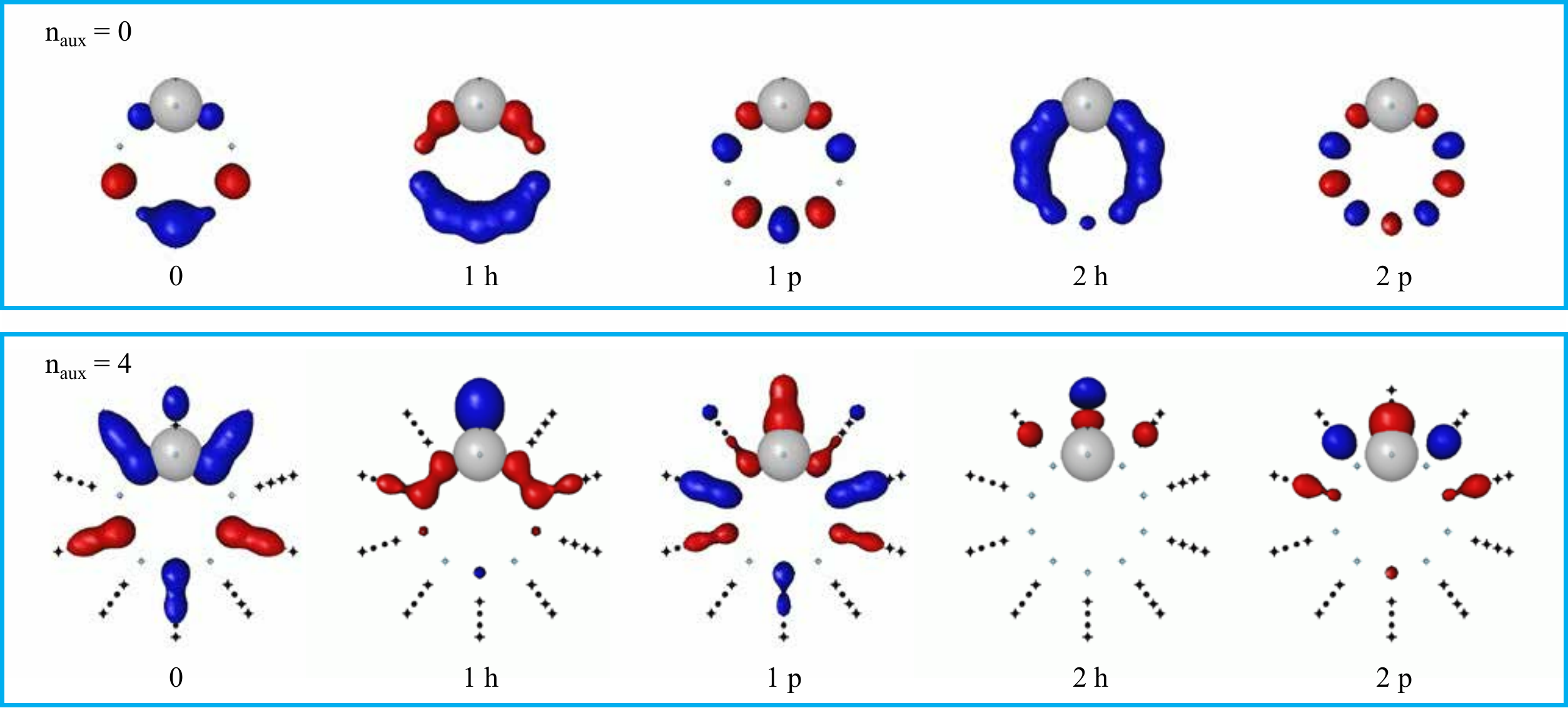}
\caption{Bath orbitals needed to describe up to the 5$^{\rm th}$ moment for the stretched H$_{10}$ ring (inner ring) at a H-H internuclear distance of 1.6\AA, as obtained without auxiliaries (upper row) and with EwDMET using 4 auxiliary orbitals (black dots in lower row) attached to each site. The fragment orbital is denoted by the grey sphere, and the bath orbitals span the physical and auxiliary (if present) space orthogonal to this fragment orbital. The labels indicate the dominant $m$ and hole/particle character of the bath states, as given by their overlap with the states of Eqs.~\ref{eqn:schmidtmom_h} and \ref{eqn:schmidtmom_p}. More details are given in the main text.} 
\label{fig:Bath_orbitals_H10}
\end{figure*}

In order to gain some physical insight into these bath states, in Fig.~\ref{fig:Bath_orbitals_H10} we plot the converged bath orbitals for a 60\% stretched H$_{10}$ ring in a minimal basis.
A (\Lowdin orthogonalized) 1s orbital is selected as a single fragment state on one of the atoms, denoted by the grey sphere, which is equivalent on all atoms. It should be noticed that a spin-restricted formalism was employed here. We also consider the cases of $n_{\rm aux}=0$ (upper row) where $\hat h_{\rm tot}=\hat f$, and of $n_{\rm aux}=4$, where four additional fictitious auxiliary sites per atom are included, and couple to their respective atom (lower row). These auxiliary sites are denoted as black dots at increased radial distances from the atom which they couple to, and are optimized such that the fragment moments of the cluster and mean-field match for each atom. The five bath orbitals plotted for each row span the fluctuation space required to faithfully match up to $n_{\rm mom}=5$. The dominant $m$ and hole/particle character of each bath orbital is also shown (though this is not unambiguous due to the orthogonalization procedure). The inclusion of the auxiliary space reduces the weight of the bath orbitals on the physical system, indicating a reduced coupling between the atoms, and an opening of the effective HOMO-LUMO gap compared to the original mean-field, to account for the qualitatively different self-consistent correlated physics. In contrast, without auxiliaries the bath states strongly resemble the original canonical occupied (hole) and virtual (particle) mean-field states, as they cannot adjust to the correlated character of the bonding. In both examples, it is also evident that in order to represent higher order moments and longer-ranged fluctuations, the corresponding bath orbitals are required to span regions more physically distant from the fragment, to allow the description of these longer path dynamics.

Before solving for the correlated EwRDMs of the cluster, care must be taken in order to avoid any double-counting of the correlation effects in the interacting cluster hamiltonian, \mbox{$\hat H_{\rm clust}$.} Indeed, the auxiliary space of the extended Fock hamiltonian is designed explicitly to modify the one-electron states to mimic the local correlation effects of each fragment region. In addition, the Coulomb and exchange terms included in the original Fock hamiltonian $\hat f$ need to be subtracted from the fragment space, since the full two-electron interaction terms are subsequently included, and would therefore double-count the Coulomb and exchange contributions of this space. The full Hamiltonian of the cluster can therefore be defined with one-electron terms, and a two-electron part which only acts over the fragment degrees of freedom, as
\begin{equation}
\hat H_{\rm clust} = {\hat h}_{\rm clust} + \frac{1}{2}\sum_{\alpha\beta\gamma\delta \in \rm frag} \left(\alpha\beta|\gamma\delta\right){\hat c}^{\dagger}_{\alpha}{\hat c}^{\dagger}_{\gamma} {\hat c}_{\delta}{\hat c}_{\beta}	. \label{eqn:clust_ham}
\end{equation}
The one-particle contribution is defined as 
\begin{equation}
\begin{split}
\hat h_{\rm clust} =& \hat P_{\rm clust} \hat h_{\rm tot} \hat P_{\rm clust}\\
                    &- \hat P_{\rm frag} \hat v^{\rm dc} \hat P_{\rm frag} - \left(\hat P_{\rm frag} \hat h_{\rm aux} \hat P_{\rm clust} + h.c.\right)\\
                    &+ \mu_{\rm bath}\hat P_{\rm bath} \hat P_{\rm bath}	,
\end{split}\label{eqn:oneelecclusth}
\end{equation}
where the exact double counting term is
\begin{equation}
v_{\alpha\beta}^{\rm dc} = \sum_{\gamma\delta \in \rm frag} \left[ \left(\alpha\beta|\gamma\delta\right) -  \left(\alpha\delta|\gamma\beta\right) \right] D_{\gamma\delta}	,
\end{equation}
corresponding to the Coulomb and exchange terms of the fragment space. In addition, the coupling between the fragment space, and any auxiliary degrees of freedom are explicitly removed in Eq.~\ref{eqn:oneelecclusth}, as these effects would also over-correlate the fragment space, whose correlation effects should only be driven by the explicit two-electron interactions. Effective correlation driven by the auxiliaries which act purely within the bath space of the cluster are however retained, ensuring that the one-electron Hamiltonian of the cluster resembles a non-interacting fragment, coupled to an bath space which is modified due to the correlated effect of the auxiliaries. Finally, to ensure that the total number of electrons in the fragment space is equal over all symmetry-equivalent sites, a chemical potential $\mu_{\rm bath}$ is optimized and applied in the bath space\cite{Wouters2016}.

In order to compare and contrast to the DMET approach, it should be pointed out that in the literature, DMET has been implemented in two different variants, often described as a `non-interacting-bath' and an `interacting-bath' formalism\cite{doi:10.1021/ct301044e,Wouters2016}. These are distinguished by whether the two-particle interaction terms of Eq.~\ref{eqn:clust_ham} are restricted to span the fragment space only, or projected into the whole cluster space. In our formulation of EwDMET, we will employ an non-interacting bath formulation, with the auxiliary hamiltonian in the bath providing the effective correlation-driven modifications to this space. The introduction of an `interacting-bath' formalism, as well as other approaches for the inclusion of non-fragment-local two-electron terms in the cluster Hamiltonian will be considered in future work, while the current `non-interacting-bath' will ensure that the large-$n_{\rm mom}$ limit of the method allows for a direct comparison to DMFT results.

The formalism above is written in a spin-orbital basis, where the density matrix, and all terms can potentially have different alpha and beta components (or even a mixed, non-collinear character). This allows for spontaneous spin-symmetry breaking in both the fragment, physical system, and auxiliary Hamiltonian. Naturally, the approach can be applied in a spin-restriced formalism by matching mean-field and correlated spin-traced moments rather the individual spin-partitioned moments. This reduces the number of auxiliary parameters as the single-particle total hamiltonian is identical for both spin channels. In section~\ref{section:results} we will discuss the results obtained with both the spin-restricted (EwDMET) and the spin-unrestricted (U-EwDMET) treatment and highlight how the increased flexibility offered by the latter improves significantly the results in the recoupling regime.

\section{Self-consistent algorithm}\label{section:algorithm}

In this section we put the components described above into a general algorithm for a EwDMET calculation in quantum chemistry. The initial choice concerns the selection of the correlated region, and the $n_{\rm frag}$ fragment orbitals to span this space. Although, there is a freedom in the exact choice of representation for such fragment orbitals, the use of atomic orbitals (AOs) seem a sensible one for a chemical system. In this work, in order to avoid dealing with a non-orthogonal basis, we employed L\"owdin-orthogonalized\cite{Loewdin1956} AOs, ensuring that fragments remain symmetry-equivalent, although other choices exists, and may be preferable. A choice of the maximum EwRDM (moment) of this fragment space to self-consistently optimize ($n_{\rm mom}$) is also made, defining the range of correlation-driven fragment fluctuations to include in the calculation. 

A self-consistent field calculation is then performed to obtain a reference Fock matrix in the basis of fragment and remaining orthonormalized environment orbitals. Couplings (${\bf \varepsilon}$) and energies (${\bf \nu }$) of $n_{\rm aux}$ auxiliaries per symmetry-equivalent fragment space are randomly initialized, according to the Hamiltonian of Eqs.~\ref{eqn:sumaux} and \ref{eqn:auxham}, and used to augment the Fock matrix of Eq.~\ref{eqn:totham}. The self-consistent coupled optimization of these quantities along with the EwRDMs is then performed, according to an iterative process with the following stages.
\begin{enumerate}[label=(\roman*)]
\item The bath orbitals required to represent the ${\tilde {\bf T}_{h/p}^{(n)}}$ hole/particle moments of ${\hat h}_{\rm tot}$ are constructed. The required $m_{\rm max}$ is found according to $\left \lceil \frac{n_{\rm mom}-1}{2} \right \rceil$, and the set of orbitals according to Eqs.~\ref{eqn:schmidtmom_h} and \ref{eqn:schmidtmom_p} constructed and orthonormalized, with any linear dependencies removed. These orbitals are combined with the fragment degrees of freedom to form a projector into the cluster space of fragment $\oplus$ bath orbitals, $P_{\rm clust}$.
\item The interacting cluster Hamiltonian is formed according to Eq.~\ref{eqn:clust_ham}.
\item By means of an accurate solver (full configuration interaction (FCI) in this work\cite{pyscf}), the cluster Hamiltonian is solved for the ground state wave function and energy.
The bath space chemical potential ($\mu_{\rm bath}$) is optimized (e.g. by bisection) to ensure that at each iteration the number of electrons in the fragment space of the correlated cluster equals the trace of the mean-field RDM over the fragment space.
Without this step, the electron number of the fragment space can fluctuate each iteration. The correlated expectation values over the fragment, ${\bf T}_{h/p}^{(n)}$, of Eqs.~\ref{eqn:CorrHole} and \ref{eqn:CorrPart} are then computed from the ground state wave function and energy of the cluster, up to the chosen maximum moment order of $n_{\rm mom}$.
\item The auxiliary parameters $\{\varepsilon;\nu\}$ and the hermitian matrix $v_c$ of the Hamiltonian of Eq.~\ref{eqn:auxham}  are optimized, by minimizing the squared error between the lattice moments (${\bf {\tilde T}}_{h/p}^{(n)}$ of Eq.~\ref{eqn:latmom_h}) and cluster moments (${\bf T}_{h/p}^{(n)}$) as
\begin{equation}
\begin{split}
C = &\sum_{\alpha \beta \in \rm frag} \sum_{z=p/h} \sum_{n=0}^{n_{\rm mom}} w_{n} \left({\tilde T}_{z, \alpha \beta}^{(n)}[v_c,\varepsilon ,\nu ]-T_{z, \alpha \beta}^{(n)}\right)^2  \\
  &+ \left(\mu_{\rm ext}[\varepsilon ,\nu ] - \mu\right)^2  ,	\label{eqn:cost}
\end{split}
\end{equation}
where $w_{n}$ are weighting parameters of the moments, which we take to be $w_{n}=\frac{1}{n!}$. This numerical minimization of $C$ is performed using analytic gradients with respect to all optimizable degrees of freedom, while ensuring that the chemical potential of this extended space matches the mean-field one of the physical space. This keeps the total number of electrons in the physical system fixed.
\item The updated auxiliary Hamiltonian parameters are used as new guess into step (i) and the self-consistency cycle is repeated until convergence in the values of the moments  ${\bf T}_{h/p}^{(n)}$ and auxiliary parameters is reached.
\end{enumerate}
It is worth noting that increasing $n_{\rm aux}$ (and therefore the dynamical resolution of the effective self-energy) only increases the computational cost of the fitting and lattice diagonalizations. It therefore does not affect the cluster size or the effort of the solver for the correlated moments, which is instead determined by the order of the highest moment to fit, and thus number of bath orbitals, as given by Table~\ref{tab:bathspaces}.

\section{Energy Functional}\label{section:energy}

In this section, we derive a local energy functional which depends purely on the static quantities self-consistently optimized in EwDMET, that is the energy-weighted reduced density matrices, or particle and hole moments of the fragment Green function, as related in section~\ref{sec:GFs}. The total energy of a system is defined as a sum of one-body and two-body contributions, as
\begin{eqnarray}
E &=&  E_{\rm nucl} + \langle {\hat T} \rangle + \langle {\hat V}\rangle \label{eqn:totE1} \\
&=&  E_{\rm nucl} + \sum_{ij} t_{ij} D_{ij} + \frac{1}{2}\sum_{ijkl} \left(il|jk\right) P_{ijkl}, \label{eqn:totE2}
\end{eqnarray}
where $E_{\rm nucl}$ is the nuclear repulsion contribution, ${\bf D}$ is the one-body density matrix, defined in Eq.~\ref{eqn:1RDM}, and ${\bf P}$ is the two-body density matrix, defined as
\begin{equation}
P_{ijkl} = \langle \Psi | c_{i}^{\dagger} c_{j}^{\dagger} c_{k} c_{l} | \Psi \rangle .\label{eqn:2RDM}
\end{equation}
Partitioning such an energy expression into local contributions connected to individual molecular fragments, $E_{\rm frag}$ is not an obvious task. These local terms should include both the trivial local one- and two-particle contributions as well as account for the interaction of the fragment with its environment. Such expression must also be size-extensive and yield the total energy upon summation over any arbitrary fragmentation of the molecule, as
\begin{equation}
E =  E_{\rm nucl} + \sum_{\lambda} E_{\rm frag(\lambda)}	. \label{eqn:Epart}
\end{equation}
The starting point for this derivation is the Galitskii-Migdal formula,\cite{Galitskii1958} which depends on the dynamical Green function, as
\begin{equation}
E = -\frac{1}{2 \pi} {\rm Tr} \left[ \int_{-\infty}^{\mu} {\rm Im}[{\bf G}(\omega)] (\omega + {\bf t}) d\omega \right]
\end{equation}
where ${\bf t}$ is the non-interacting Hamiltonian of the system, including kinetic and nuclear potential terms. This energy expression is connected to the 0$^{\rm th}$ and $1^{\rm st}$ moments of the hole spectral function as expressed in Eqs.~\ref{eqn:Spectrum} and \ref{eq:mom_distr_h}, to give
\begin{equation}
    E = \frac{1}{2} {\rm Tr} \left[  {\bf T}_{h}^{(0)}  {\bf t} \right] + \frac{1}{2} {\rm Tr} \left[  {\bf T}_{h}^{(1)} \right] . \label{eqn:GM_formula_mom}
\end{equation}
When tracing the above formula over a local (e.g. fragment) Green function, the energy functional of Eq.~\ref{eqn:GM_formula_mom} refers to the energy contribution of the corresponding local fragment, with the desired additive property of Eq.~\ref{eqn:Epart}. In other words, the total energy per fragment region ($E_{\rm frag}$) can be exactly calculated by tracing over the fragment zeroth and first hole moments,\cite{Nooijen1992,Nooijen1993} which are self-consistently optimized through the calculation. Since in EwDMET the bath space required to represent $n_{\rm mom}=1$ is identical to $n_{\rm mom}=0$ (see Table~\ref{tab:bathspaces}), this means that this energy can be evaluated self-consistently with {\em no} additional computational effort.

The correlated physics contained in higher order moments appears explicitly here, since despite being a one-particle quantity, ${\bf T}_{h}^{(1)}$ contains the two-particle contributions required for the total energy within it. This can be be seen by expanding Eq.~\ref{eqn:mom_h_commutators} for $n=1$ and by using the definition of the cluster Hamiltonian given in Eq.~\ref{eqn:clust_ham}, as
\begin{equation}
T_{h, \alpha \beta}^{(1)} = \bra \Psi| \hat c_{\alpha}^\dagger [\hat c_{\beta}, \hat H_{\rm clust}] |\Psi \ket	.
\end{equation}
By expanding this commutator as
\begin{equation}
\begin{split}
[\hat c_{\beta}, \hat H_{\rm clust}] =& \sum_{ij \in \rm clust} h^{\rm clust}_{ij} \left( {\hat c}_{\beta} {\hat c}^{\dagger}_{i} {\hat c}_{j} - {\hat c}^{\dagger}_{i} {\hat c}_{j}{\hat c}_{\beta} \right) \\ 
                 +\frac{1}{2}&\sum_{ijkl \in \rm frag} \left(ij|kl\right) \left( {\hat c}_{\beta} {\hat c}^{\dagger}_{i}{\hat c}^{\dagger}_{k} {\hat c}_{l}{\hat c}_{j} - {\hat c}^{\dagger}_{i}{\hat c}^{\dagger}_{k} {\hat c}_{l}{\hat c}_{j}{\hat c}_{\beta}\right)
\end{split}
\end{equation}
\begin{eqnarray}
                 &=& \sum_{j \in \rm clust} h^{\rm clust}_{\beta j} {\hat c}_{j} + \sum_{jkl \in \rm frag} \left(\beta j|kl\right) {\hat c}^{\dagger}_{k} {\hat c}_{l}{\hat c}_{j} ,
\end{eqnarray}
the first hole moment can then written as
\begin{equation}
    T^{(1)}_{h, \alpha\beta} = \sum_{j \in \rm clust} h^{\rm clust}_{\beta j} D_{\alpha j} + \sum_{jkl \in \rm frag} \left(\beta j|kl\right) P_{\alpha k l j}	,	\label{eqn:Hole1Ex}
\end{equation}
The explicit dependence of the first hole moment on the two-body density matrix is evident in Eq.~\ref{eqn:Hole1Ex}, where by comparing to Eq.~\ref{eqn:totE2}, it can be seen that the second term yields twice the two-particle contribution to the fragment energy upon tracing over the fragment space, as
\begin{equation}
\sum_{\alpha jkl \in \rm frag} \left(\alpha j|kl\right) P_{\alpha k l j} = 2\langle {\hat V} \rangle_{\rm frag}	,
\end{equation}
while the first term of Eq.~\ref{eqn:Hole1Ex} contains one-electron, Coulomb and exchange contribution.
By substituting Eq.~\ref{eqn:Hole1Ex} into Eq.~\ref{eqn:GM_formula_mom}, we obtain the final expression for the fragment energy, as
\begin{equation}
    E_{\rm frag} = \sum_{\substack{\alpha {\in \rm frag} \\ j {\in \rm clust}}} \frac{t_{\alpha j} + h^{\rm clust}_{\alpha j}}{2} D_{\alpha j} + \frac{1}{2} \sum_{\alpha jkl \in \rm frag} \left(\alpha j|kl\right) P_{\alpha k l j} . \label{eq:energy_final}
\end{equation}
which includes the one-body density matrix with one index over the entire cluster space (including bath), rather than just the fragment degrees of freedom.
This accounts for the (optimized) description of the fragment-bath coupling which correctly includes Coulomb and exchange contributions via the effective one-particle operator, $\frac{1}{2}\left(t + h^{\rm clust}\right)$, while the double-counting of these contributions within the fragment exactly cancels.

At convergence the mean-field (${\bf {\tilde T}}_h^{(n)}$) and correlated (${\bf T}_h^{(n)}$) moments will match over the fragment space, however they will not in general over the whole cluster. Therefore, the one-body contribution to Eq.~\ref{eq:energy_final} will not be identical if projected from the converged mean-field full system RDM or from the correlated cluster RDM. However, the latter is the appropriate choice, since it matches the one-body contribution from the first-order hole moment as given in Eq.~\ref{eqn:Hole1Ex}. The fragment energy expression of Eq.~\ref{eq:energy_final} is equivalent to the one employed in the DMET method previously\cite{Wouters2016}, which is now derived from the Galitskii-Migdal energy functional and justified above. However, in contrast to DMET, in the EwDMET approach the two-particle contribution is also self-consistently optimized as it is a contribution from the first-order hole moment, as opposed to just the self-consistent one-body density matrix contribution to the fragment energy in DMET.

\section{Results: Hydrogen rings}\label{section:results}

\begin{figure*}[th!]
\includegraphics[scale=0.5,trim={0 0 0 0},clip]{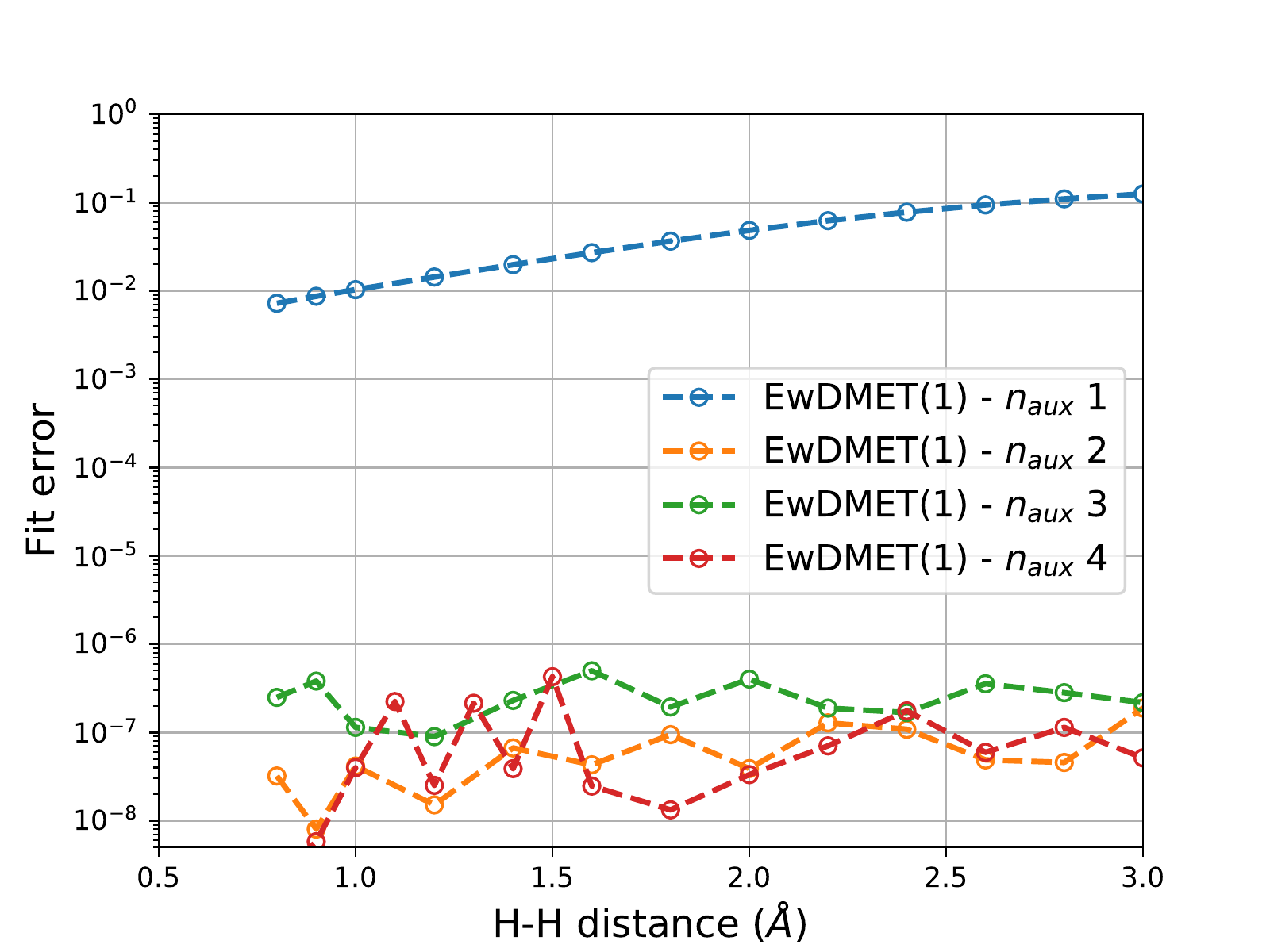}
\includegraphics[scale=0.5,trim={0 0 0 0},clip]{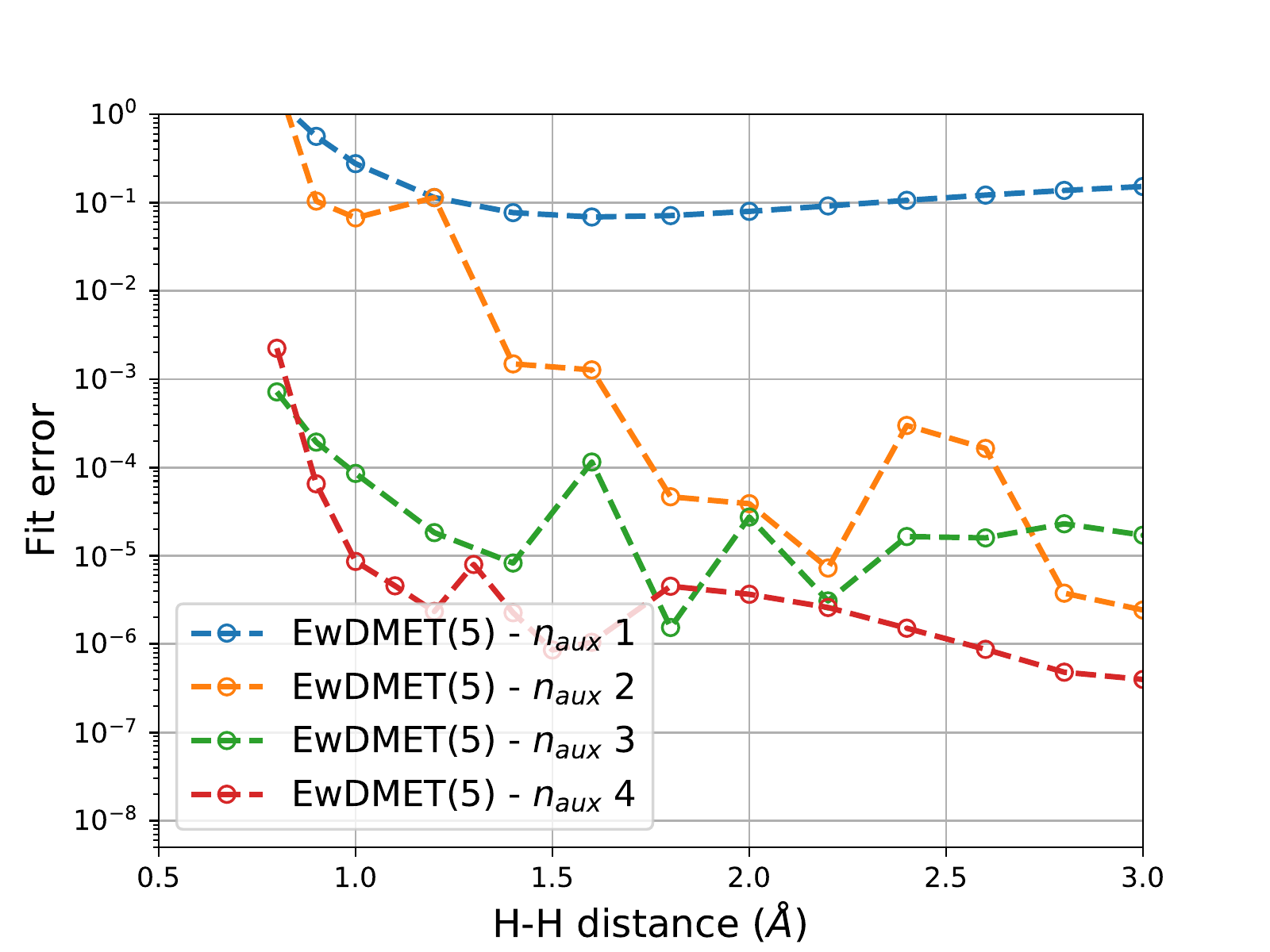}
\caption{
EwDMET fit error for the dissociation of H$_{10}$ ring calculated with different numbers of fit auxiliary degrees of freedom per fragment space, for the self-consistent optimization with $n_{\rm mom}=1$ (left panel) and $n_{\rm mom}=5$ (right panel).}
\label{fig:H10_result_fit_error}
\end{figure*}

The EwDMET method has previously been applied to Hubbard models of correlated materials,\cite{Fertitta2018} where increasing the number of moments optimized in the method demonstrated improvability of results and allowed for correlation driven phase transitions to emerge, which were missed at low order. In order to provide a stern yet controllable test as to the reliability of EwDMET across different correlation regimes for {\em ab initio} chemical systems, we consider here the symmetric dissociation of hydrogen rings into individual hydrogen atoms. This model system has provided a benchmark for many different methods, with a large quantity of comparison data available\cite{DMRG_ref,DMFT_ref,doi:10.1021/ct301044e,PhysRevX.7.031059}. While these systems are superficially related to the physics of the half-filled one-dimensional Hubbard chains, the introduction of true, long-ranged Coulomb potentials introduces much more subtle and rich physics, which is still to be fully understood in the large chain limit\cite{PhysRevX.7.031059}.

In the equilibrium geometry the wave function is dominated by its mean-field character, though a reasonable amount of correlation still exists in this regime. Upon stretching, strong spin fluctuations dominate and a high-level of theory is required to obtain a qualitatively correct description, as spurious ionic configurations in the restricted mean-field reference result in a vastly too high energy and resulting dissociation error. This necessitates a multireference treatment of the wave function if building on this mean-field picture, and a complete breakdown of any finite-order perturbation theory. In the infinite chain limit, there are a formally infinite number of these strong correlation centers, which results in an entangled, and infinitely degenerate state in the restricted basis. 

Analogues to formal phase transitions are found, as the description of the system beyond some bond length will want to spontaneously break spin symmetry. This can be described as a metal to Mott insulator transition, with the Coulomb repulsion eventually dominating, and a significant single-particle gap opening with respect to electron number change. While the mean-field picture will reduce the HOMO-LUMO gap to zero in the restricted basis, the true gap will tend to the sum of the first ionization potential and electron affinity of a Hydrogen atom.
To remedy this error, the EwDMET method in this limit will ensure the hybridization of the physical atomic states with the auxiliary system will yield an effective enlarged mean-field gap in the physical system. This effective gap does not correspond to the HOMO-LUMO gap of the total hamiltonian $\hat h_{\rm tot}$ which spans all degrees of freedom, but it is rather calculated from the projection of the total mean-field spectrum into the physical system.
Furthermore, the one-dimensional nature of this system allows for numerically exact results from DMRG to be compared to, as well as benchmarking the method against both DMET and DMFT results for this system. In contrast to Ref.~\onlinecite{doi:10.1021/ct301044e}, the DMET results in this section are obtained in the non-interacting-bath formalism to allow a clearer comparison to the EwDMET results, in which case they correspond exactly to EwDMET in the $n_{\rm mom}=n_{\rm aux}=0$ limit.

The convergence of both the number of auxiliary states, as well as the number of moments in the fragment description in order to achieve the correct physics of this system will be analysed in this section. To denote the order of the self-consistent moments in each calculation, this will be put in parentheses, i.e. EwDMET(5) denotes an $n_{\rm mom}=5$ self-consistent calculation. All calculations presented here were performed using a minimal Gaussian basis set\cite{STO-nG_basis_ref}, with STO-3G used for the H$_{10}$ ring and STO-6G for H$_{50}$ to compare it with pre-existing benchmarks\cite{DMRG_ref,DMFT_ref}. The fragment space is composed of a single 1s-like atomic orbital obtained by \Lowdin orthonormalization of the atomic basis set, to ensure that translational symmetry of the fragment space is maintained, though other (perhaps preferable) choices may exist. The H$_{10}$ system is small enough to allow exact diagonalization (FCI) results on the full system to compare, while in the H$_{50}$ system, DMRG reference values are available\cite{DMRG_ref}. 

\subsection{Convergence of auxiliary space}

We first consider the convergence in number of auxiliary degrees of freedom required ($n_{\rm aux}$) in order to match the mean-field and correlated moments, across different bond lengths and moment expansions. This is shown in Fig.~\ref{fig:H10_result_fit_error}, where the convergence of the squared error in the moments (the cost function of Eq.~\ref{eqn:cost}) is shown for all bond lengths, and for differing number of auxiliary orbitals per fragment. The left panel shows the convergence for $n_{\rm mom}=1$, while the right panel shows the convergence for $n_{\rm mom}=5$. Since the absolute values of the moments increase substantially with order, the absolute values of the fit error are difficult to compare between different $n_{\rm mom}$.

The trend of Fig.~\ref{fig:H10_result_fit_error} is certainly that the fit error generally reduces with increasing number of auxiliary sites per fragment. This is to be expected, as increased flexibility is afforded in Eqs.~\ref{eqn:latmom_h} and \ref{eqn:latmom_p} with this increase, and this allows the total system to better reproduce the local, correlated moments of the cluster. However, as $n_{\rm aux}$ increases, it also changes the bath orbitals of the cluster, as well as the Hamiltonian used in this bath space, as it also changes the effective mean-field coupling between the fragments. This overall self-consistency is far easier and more stable at low order, where it can be seen that two auxiliaries is enough to match the zeroth and first moments at all bond lengths up to numerical precision.

As the number of moments to self-consistently optimize increases, the number of auxiliaries required to fulfil this criteria is also seen to increase, as demonstrated by the right panel of Fig.~\ref{fig:H10_result_fit_error}. Somewhat counter-intuitively, we also find that the convergence is slower and more difficult at weaker correlation strengths. This is likely a consequence of the fact that the changes to the eigenvalue spectrum are more dispersive in terms of the energy range of these changes, and also the fact that there are likely many local minima to the optimization of the non-convex function of Eq.~\ref{eqn:cost}. This was also found to be the case in the low-$U$, metallic regime of the Hubbard model results of Ref.~\onlinecite{Fertitta2018}, and the improvement of the self-consistent procedure in this regime is a source of further active research. However, as $n_{\rm aux}=4$ yields a decent fit error for all bond lengths up to $n_{\rm mom}=5$, we use this as the default value for all subsequent calculations.


\subsection{Dependence on order of self-consistent moments}

\begin{figure}[]
\includegraphics[scale=0.5,trim={0 0 0 0},clip]{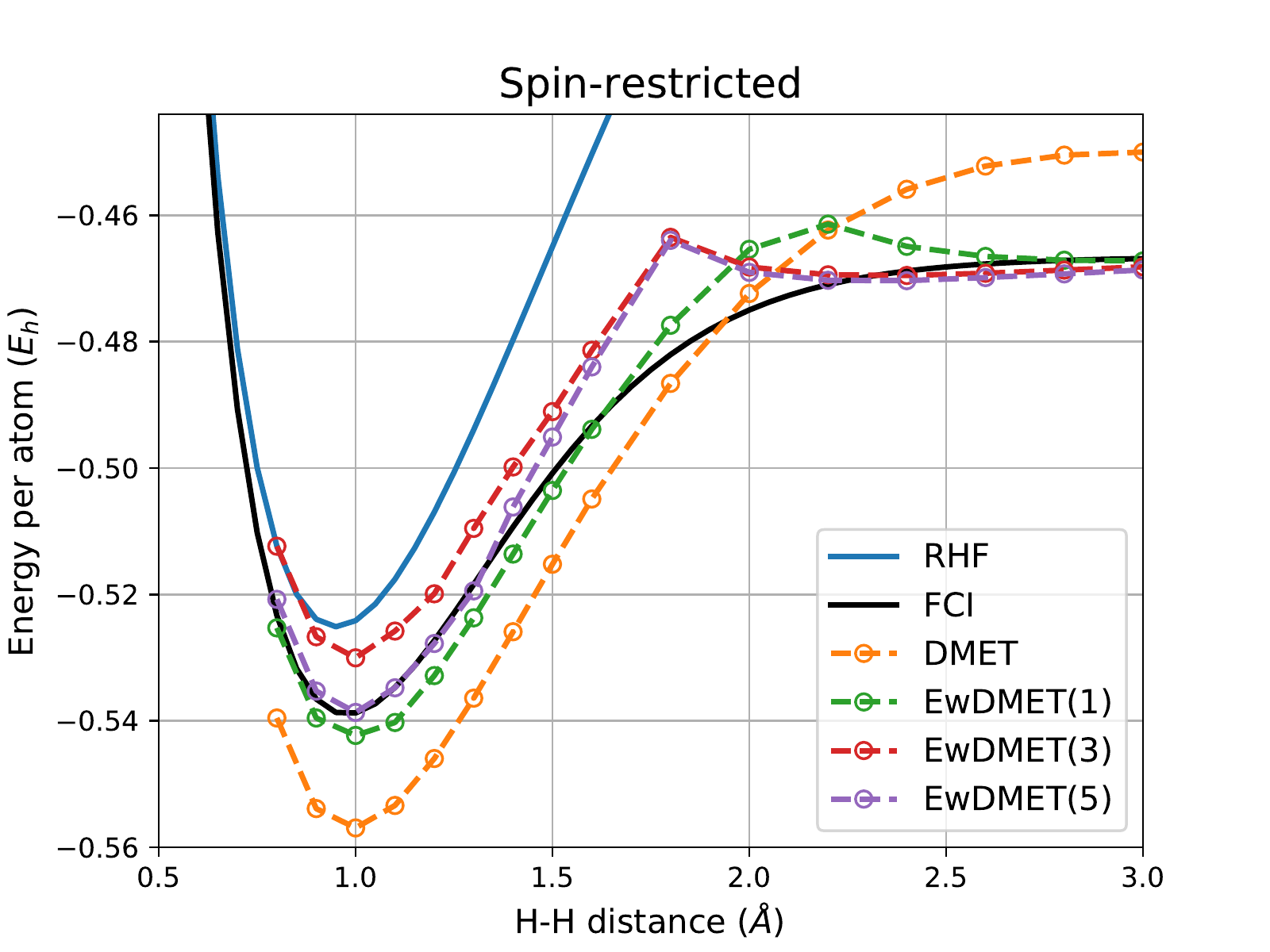}	\\
\caption{Binding curve for the symmetric stretch of the STO-3G H$_{10}$ ring calculated using EwDMET in a restricted formalism, with a single atom fragment space. Four auxiliary states per fragment were optimized, up to a maximum of the $1^{\rm st}$, $3^{\rm rd}$ and $5^{\rm th}$ moment fluctuations. Comparison to FCI, RHF and DMET in its non-interacting bath formalism are also reported.}
\label{fig:H10_result_RHF}
\end{figure}

\begin{figure}[h]
\includegraphics[scale=0.5,trim={0 0 0 0},clip]{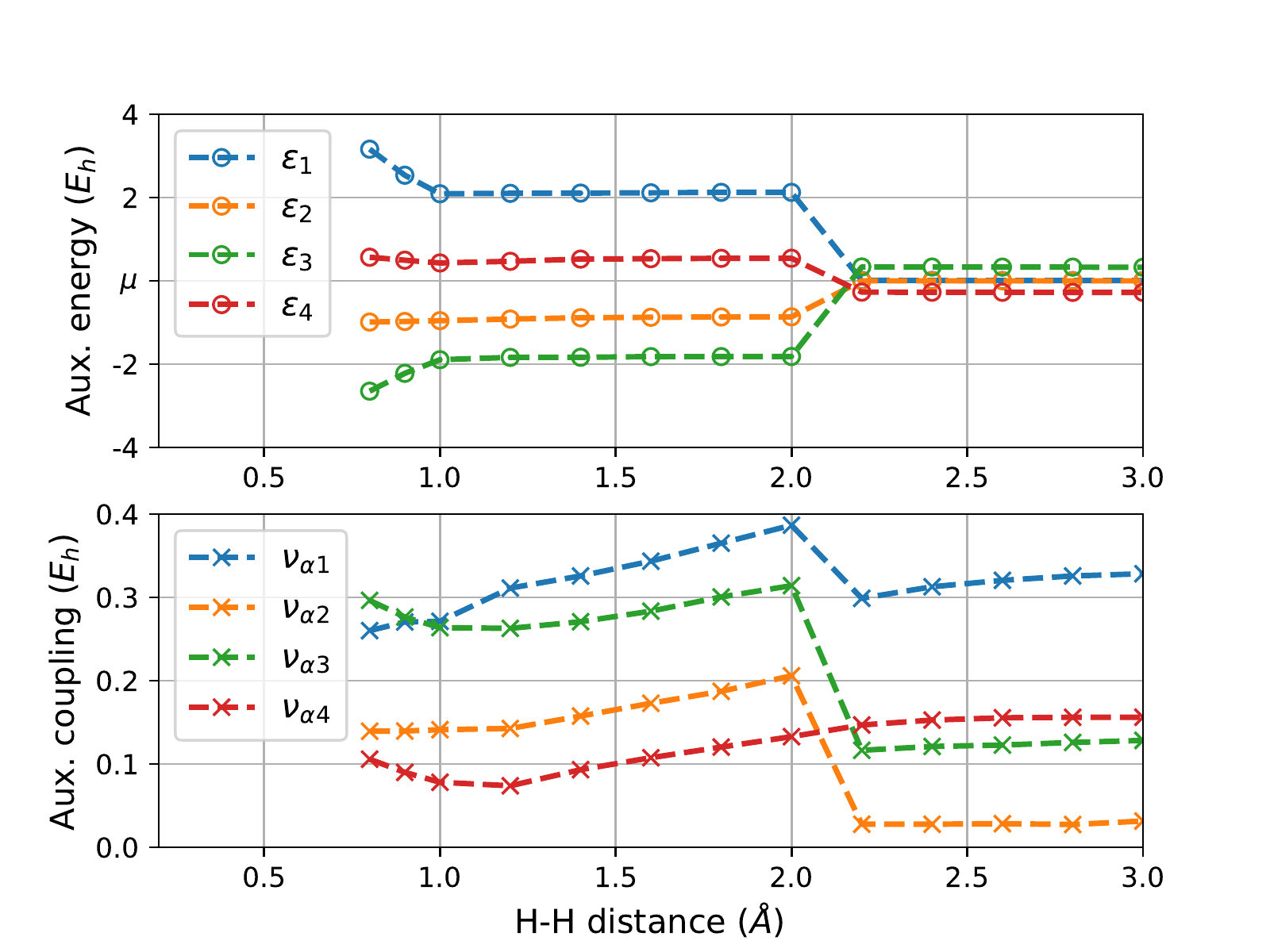}
\caption{Self-consistent auxiliary energies and couplings for the spin-restricted H$_{10}$ ring as a function of the internuclear distance. EwDMET results were obtained using 4 auxiliaries per site and fitting up to the $1^{\rm st}$-order moments.}
\label{fig:H10_result_params}
\end{figure}

The finer description of longer-ranged correlated changes to the fragment propagator (or `dynamical' correlation in the self-energy rather than traditional chemists sense of the term) obtained by increasing the auxiliary space and the order of the moments, is also reflected in the values for static quantities such as the total energy of the system. In Fig.~\ref{fig:H10_result_RHF} we report the binding energy of H$_{10}$ calculated with EwDMET in an RHF basis through different order moments ($n_{\rm mom}$). For this system, it is possible to compare to FCI calculations, and we include converged energies for $n_{\rm mom}=1, 3, 5$, while also comparing to DMET, where $n_{\rm mom}=n_{\rm aux}=0$. The addition of the auxiliary space is able to correctly describe the dissociation regime, even in this restricted (paramagnetic) Hartree-Fock basis. Insight into the system can also be found from analysis of the parameters in this auxiliary system, as shown in Fig.~\ref{fig:H10_result_params}, with both the energy of the converged auxiliaries, and strength of their coupling to the physical fragment atom given. 

Within the equilibrium regime, the energies are relatively widely spaced, with the couplings increasing upon stretching, reflecting the increased self-energy and strong quantum fluctuations in the system. At 2.2\AA, the transition to the correlated quasi-Mott insulator state is reflected in the collapse of energy of the poles to be around the chemical potential. These auxiliary states hybridize with the large degenerate set of states in this dissociated limit at the same energy, and hence allow for a gap to open in the single-particle spectrum in order to obtain the quantitatively correct physics. In contrast, the DMET approach is not able to benefit from the role of this auxiliary space, and therefore is qualitatively wrong in this regime, although at very large bond distance, it does eventually tend to the correct asymptote.

While increasing the order of the self-consistent moments improves the results, this improvement is not necessarily either variational or monotonic compared to FCI results in Fig.~\ref{fig:H10_result_RHF}. While the $n_{\rm mom}=0$ (DMET) result is particularly bad at equilibrium, where it captures over twice the correlation energy it should, the $n_{\rm mom}=1$ result appears much better than the $n_{\rm mom}=3$ energy at this geometry. However, it is found that the $n_{\rm mom}=5$ result in this equilibrium region is in excellent quantitative agreement with FCI. Despite this, in the intermediate `recoupling' region, where there is a strong competition between one-body and Coulomb effects, $n_{\rm mom}=3$ and $5$ values are particularly erroneous. This is due to a sharp change in the character of the state, and two competing solutions at this point leading to a cusp in the energy profile. Even for the $n_{\rm mom}=1$ results, while the sharp transition is smoothed, it is still particularly in error at this point. However, it is worth stressing that for all $n_{\rm mom}$, the only knowledge of the explicit Coulomb interactions in this approach is the value of the single $(1s 1s | 1s 1s)$ integral in Eq.~\ref{eqn:clust_ham}, where all local $1s$ orbitals are on the same atomic site, and so quantitative agreement with FCI in any regime seems impressive in that context.

\begin{figure}[]
\includegraphics[scale=0.5,trim={0 0 0 0},clip]{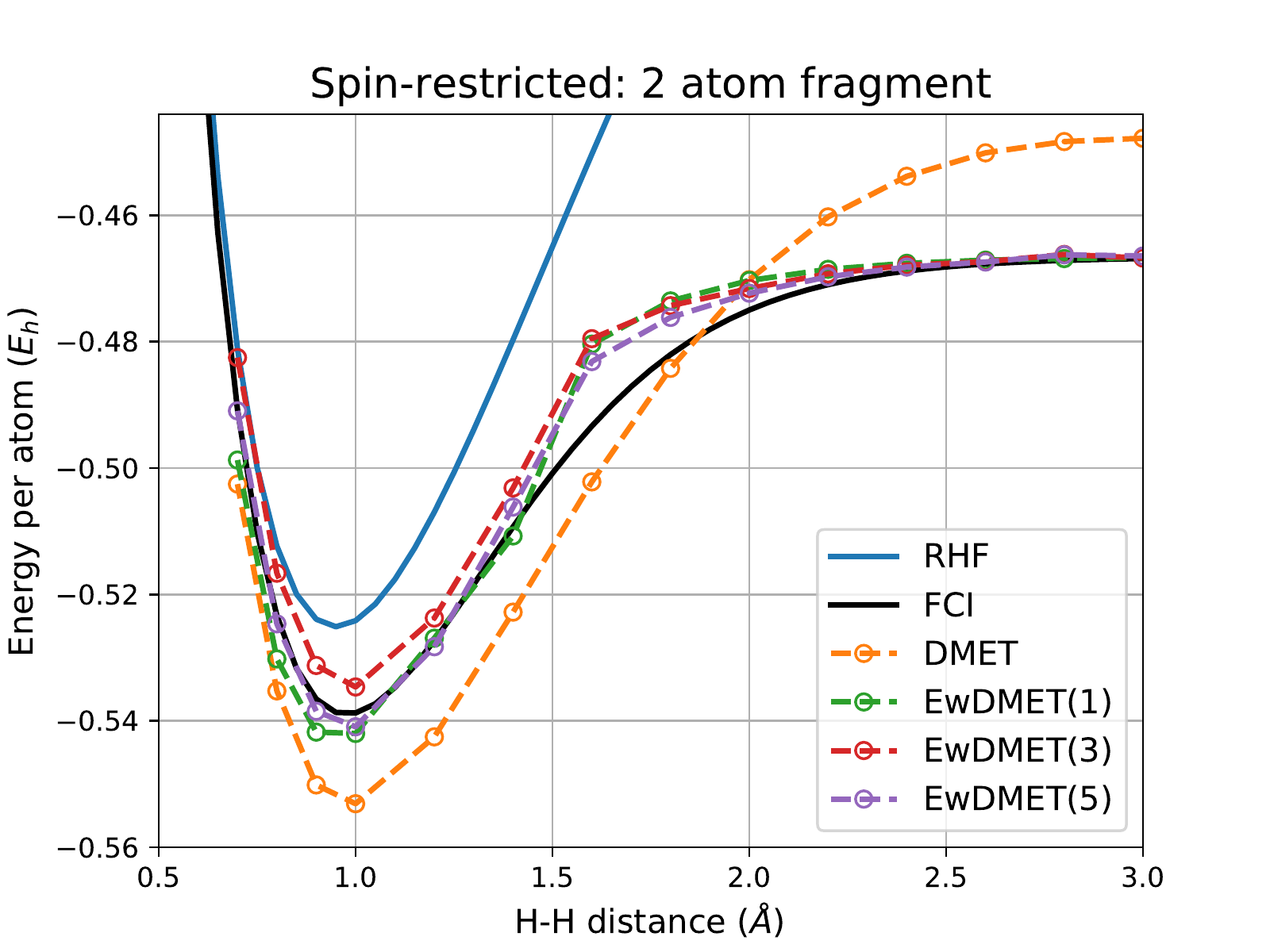}	
\caption{Binding curve for the symmetric stretch of the STO-3G H$_{10}$ ring calculated using EwDMET in a restricted formalism, with a two-atom fragment space. Four auxiliary states per fragment were optimized, up to a maximum of the $1^{\rm st}$, $3^{\rm rd}$ and $5^{\rm th}$ moment fluctuations. Comparison to FCI, RHF/UHF and DMET/U-DMET in its non-interacting bath formalism are also reported.}
\label{fig:H10_2-frag_RHF}
\end{figure}

We consider two approaches to improve on this description. Firstly, a systematic expansion to exactness of the method can be achieved via an enlargement of the number of orbitals in the fragment space, which can be straightforwardly included within the scheme. This also enlarges the number of fragment and bath degrees of freedom that need to be included in the cluster, however, it then allows for an explicit inclusion of nearest atom, non-local 2-body interactions in the cluster Hamiltonian of Eq.~\ref{eqn:clust_ham}. The results are shown in Fig.~\ref{fig:H10_2-frag_RHF}. In this case, the DMET results are still computed with $n_{\rm aux}=0$, and the correlation potential develops different diagonal elements to explicitly break the symmetry between the two atoms within the fragment, in order to match the density matrices. However, despite the inclusion of the inter-atom two-electron Coulomb terms in the cluster Hamiltonian, the DMET results for the 2-atom fragment are almost identical to the single atom fragment, with no qualitative improvement in the results. This contrast with the EwDMET results which are improved relative to FCI at nearly all bond lengths. In addition, the variation between the the different $n_{\rm mom}$ values is reduced, while the `cusp' feature in the energy profile is also substantially ameliorated. Furthermore, the correlation potential does not break the symmetry within the fragment space at any bond lengths, while this quantity tends to zero at dissociation, with the auxiliary space assuming all responsibility for the matching of the moments, without the need to break symmetry within the fragment at all. By way of comparison of computational effort, the cluster dimensionality for single-atom fragment EwDMET(3) is the same as two-atom fragment DMET, as well as two-atom fragment EwDMET(1) (four orbitals). Clearly, at least some inclusion of higher moment fluctuations is extremely beneficial for the description of the system which is not captured via simple enlargement of the fragment size.


\begin{figure}[]
\includegraphics[scale=0.5,trim={0 0 0 0},clip]{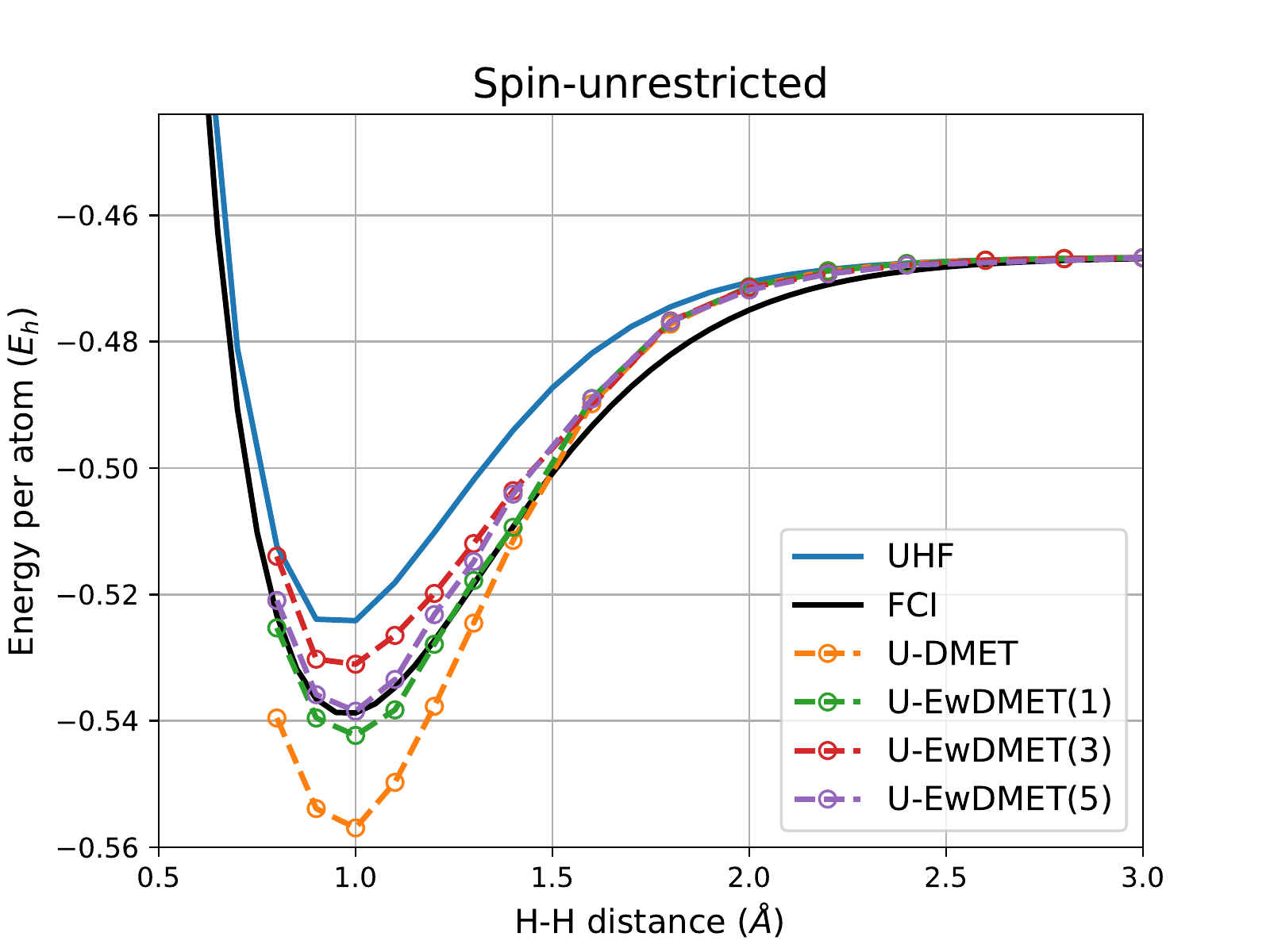}
\caption{Binding curve for the symmetric stretch of the STO-3G H$_{10}$ ring calculated using EwDMET in an unrestricted formalism, with a single atom fragment space. Four auxiliary states per fragment were optimized, up to a maximum of the $1^{\rm st}$, $3^{\rm rd}$ and $5^{\rm th}$ moment fluctuations. Comparison to FCI, UHF and U-DMET in its non-interacting bath formalism are also reported.}
\label{fig:H10_result_UHF}
\end{figure}

An alternate approach to improve the results is to allow the spin-symmetry of the fragment to break, thereby allowing explicit spin-magnetization order to develop along the chain spontaneously. This can be achieved by moving to an {\em unrestricted} formulation of the self-consistency, and in the optimization of the auxiliary space. Returning to a single-atom fragment space, these results are shown in Fig.~\ref{fig:H10_result_UHF}. In the equilibrium region, the results are essentially identical to the restricted formalism, and spin-symmetry is maintained. However, beyond $1.1$\AA~ the spin-symmetry begins to break, giving a smooth crossover in the energetics to the spin-broken solution. At all bond lengths, the spin is slightly less broken than the corresponding UHF solution, but grows towards dissociation, where it reaches an entirely spin-polarized fragment solution. At this point, the order of the moment expansion also becomes largely irrelevant, as the correlated physics is now obtained at the one-particle level through the symmetry-breaking. The auxiliary space also decouples from the physical system, as the simple UHF solution is obtained in this limit. This ensures that the unrestricted DMET results also have this correct limit. However, the unrestricted-EwDMET(5) results now provide an excellent quantitative description of the energetics across the entire binding.

\begin{figure}[]
\includegraphics[scale=0.5,trim={0 0 0 0},clip]{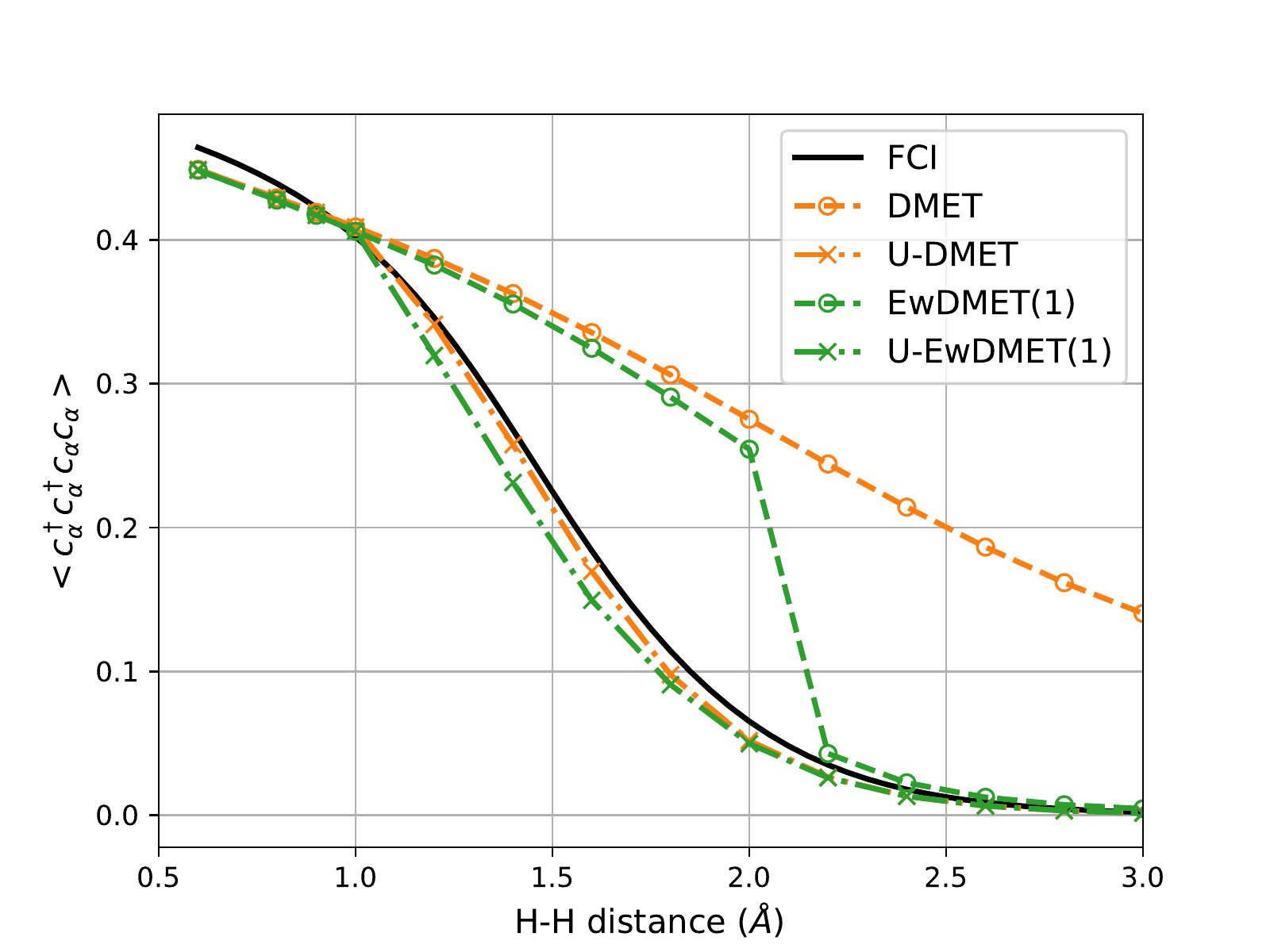}
\caption{Double occupancy of the fragment atom for STO-3G H$_{10}$ ring as a function of the internuclear distances. EwDMET and U-EwDMET results were obtained using 4 auxiliaries per site and fitting up to the $1^{\rm st}$ moment. FCI, and standard (U-)DMET are reported as comparison.}
\label{fig:H10_result_doubleocc}
\end{figure}

We can also move beyond a solely energetic description of this system, by considering the local two-particle density matrix of the fragment. The diagonal element of this in the fragment is computed, and compared to exact FCI results in Fig.~\ref{fig:H10_result_doubleocc}, in both a restricted and unrestricted formulation for $n_{\rm mom}=1$ and $n_{\rm mom}=0$ (DMET). The restricted formulation for the fragment-environment coupling induces a sharp transition at intermediate bonding (even though the energy profile for $n_{\rm mom}=1$ is smooth), denoting this crossover between the competing states. The restricted $n_{\rm mom}=0$ results are continuous, however substantially in error in all regimes due to the lack of important non-local spin fluctuations. Higher-order moment optimization in the restricted case (not shown) moves the transition point to lower bond lengths, but does not remove this transition entirely. However, the double occupation of the fragment in the spin-broken case is quantitatively correct across the range of geometries, with only relatively weak dependence on $n_{\rm mom}$. This is because the strong spin fluctuations at dissociation are now suppressed, with the spin now frozen and captured at the (static) mean-field level.

\begin{figure}[]
\includegraphics[scale=0.5,trim={0 0 0 0},clip]{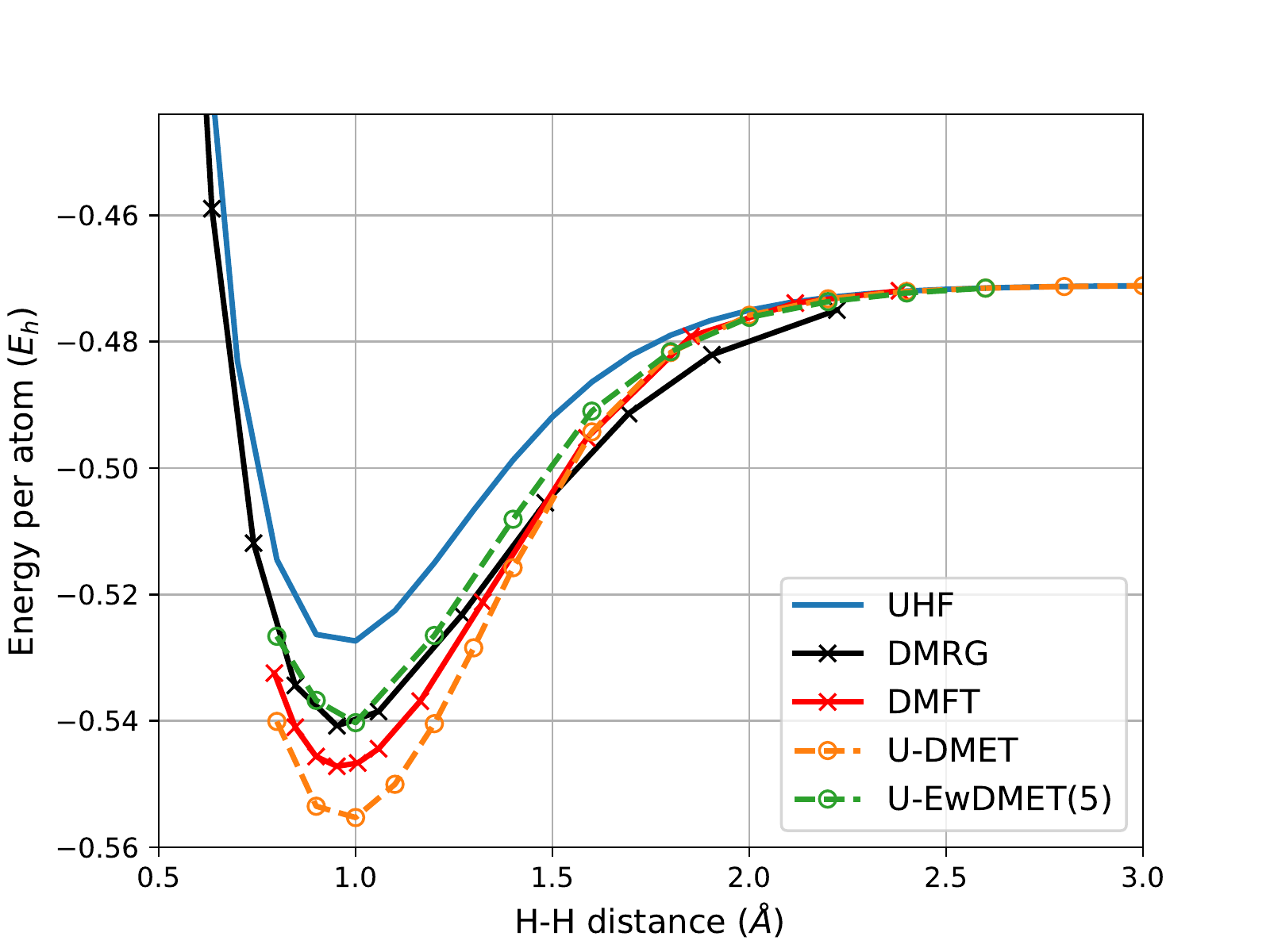}
\caption{Binding curve of an H$_{50}$ ring calculated using U-EwDMET with 4 auxiliaries per site, and self-consistent up to the $5^{th}$-order moment. Comparison to DMRG \cite{DMRG_ref}, DMFT \cite{DMFT_ref} UHF and U-DMET are also reported. Both U-EwDMET and U-DMET results were obtained using a single impurity orbital.  The basis set STO-6G was employed for all these calculations.}
\label{fig:H50_result_UHF}
\end{figure}

Due to the local nature of this approach, the scaling of the EwDMET method with system size is only mean-field and it is simple to extend the full system to much larger sizes. We therefore move to an unrestricted study of the H$_{50}$ ring, which in an STO-6G basis has been able to be numerically solved with the DMRG to provide benchmark values\cite{DMRG_ref}. Due to the similar embedding environment to the H$_{10}$ system, we initialize the auxiliary system at the parameters which converged the smaller ring, which are then relaxed to account for the longer-ranged effects. The result for the symmetric binding profile within unrestricted EwDMET up to $n_{\rm mom}=5$ are shown in Fig.~\ref{fig:H50_result_UHF}. Perhaps unsurprisingly the results are of a very similar quality to that of the H$_{10}$ system, where they are in excellent agreement with the reference DMRG at dissociated geometries, and around equilibrium (where the higher order fluctuations are seen to be important), with only a slightly spuriously high energy in the almost dissociated limit, where non-local correlation effects are important and neglected in the method. 

It is also possible to directly compare to DMFT results for the same system, which in Ref.~\onlinecite{DMFT_ref} were also performed in the same basis and within an unrestricted treatment. As discussed in Sec.~\ref{sec:intro}, the physics and properties of EwDMET and DMFT should formally match in the limit of $\{ n_{\rm aux}, n_{\rm mom} \} \rightarrow \infty$, and so this allows for comparison between these methods for {\em ab initio} systems. The methods match exactly in the dissociated and near-dissociated limit, where both tend too quickly to the atomic limit due to the lack of non-local correlation in both methods. However, at equilibrium geometries the DMFT results are significantly lower, overbinding the system. There are a number of possible reasons for the discrepancy in this regime. Firstly, it could be that $n_{\rm mom}=5$ is not sufficient to be in the large-moment limit, where quantum fluctuations are captured on all length (or time) scales within the EwDMET approach, and where agreement with DMFT should be found. This would mean that the results at $n_{\rm mom}=5$ are fortuitously good, benefiting from cancellation of errors between the neglect of non-local two-body interactions and long-range one-body correlated quantum fluctuations. The second possibility has to do with approximations in the numerical procedures within both methods. These include the numerical fitting of the hybridization function in DMFT, or the auxiliary system in EwDMET, or even the choice of orthogonalization in order to define the fragment orbital. Finally, it is also possible that the differing energy functionals between the DMFT and EwDMET methods could contribute to a discrepancy in the values obtained. All of these possibilities will be investigated in future work.

\section{Conclusion}

We have presented an approach to embed open correlated molecular fragments within a wider extended mean-field system, which allows for a systematically improvable description in the range of the correlation-driven quantum fluctuations into this environment. These fluctuations are characterized by the `moments' of the particle and hole distributions of the Green function of the fragment, which can be cast as a series of static energy-weighted one-body density matrices. Rigorous mapping of the fragment to a local ground-state quantum problem is presented, along with an approach to achieve self-consistency of these fluctuations via the optimization of a fictitious system of auxiliary states which couple to the correlated fragment. We present an energy functional for these local fragments relating to the Migdal-Galitskii form. Application to the symmetric dissociation of hydrogen atoms presents a controllable initial system where long-range realistic Coulomb interactions are present, and where rigorous benchmark values are known in large system limits to compare against. 

Within this framework, the importance of describing these longer-ranged fluctuations is shown, as well as the convergence in the size of the auxiliary system to achieve self-consistency in these fluctuations. Furthermore, the ability to describe the system within a spin-symmetry-broken mean-field environment was demonstrated, which is essential within the single atomic fragment embedding for achieving smooth and quantitatively accurate results. These accurate results were obtained despite only one explicit two-electron term in the full system Hamiltonian considered. Despite these successes, challenges remain and are a subject of ongoing research. These include the removal of the numerical fitting step, which has difficulties in the limit of high orders of $n_{\rm mom}$ and weak correlation regimes, to be replaced with a more robust analytic approach which will be presented in a forthcoming publication. The method should also be tested on more complex and subtle correlated systems, where many degrees of freedom must be embedded, and convergence of non-local correlation effects within large basis sets is required.

\section{Acknowledgements}

G.H.B. gratefully acknowledges support from the Royal Society via a University Research Fellowship, in addition to funding from the European Union's Horizon 2020 research and innovation programme under grant agreement No. 759063. We are also grateful to the UK Materials and Molecular Modelling Hub for computational resources, which is partially funded by EPSRC (EP/P020194/1).

%

\end{document}